\documentclass{article}

\RequirePackage{amsthm,amsmath,amsfonts,amssymb}
\RequirePackage{natbib}
\RequirePackage{graphicx}
\usepackage{url}
\usepackage{tikz, pgfplots}
\pgfplotsset{compat=1.18}
\usetikzlibrary{positioning}
\usepackage{booktabs}
\usepackage{authblk}
\usepackage[margin=3cm, top=2.5cm]{geometry}

\begin{document}

\title{A capture-recapture hidden Markov model framework for register-based inference of population size and dynamics}
 
\author[1]{Lucy Y. Brown\thanks{Corresponding Author: lyb3@kent.ac.uk}}
\author[2]{Eleni Matechou}
\author[3]{Bruno Santos}
\author[4,5]{Eleonora Mussino}
 
\affil[1]{School of Engineering, Mathematics and Physics, University of Kent, UK}
\affil[2]{School of Mathematical Sciences, Queen Mary University of London, UK}
\affil[3]{CEAUL - Centro de Estatística e Aplicações, Faculdade de Ciências, Universidade de Lisboa, Lisbon, Portugal}
\affil[4]{Department of Sociology, Stockholm University, Sweden}
\affil[5]{Department of Sociology, Umeå University, Sweden}
 
\date{}
 
\maketitle

\begin{abstract}
Accurate inference on population dynamics, such as migration and changes in population size, is essential for policymaking, resource allocation and demographic research. Traditional censuses are expensive, infrequent and not timely, leading many countries to adopt administration register-based approaches to replace or complement them. A primary challenge in this shift is that such registers are incomplete: even when individuals are present in the population, their activities may not generate records in specific registers in a given period, resulting in false negative observation error at the register level. Conversely, some registers do not constitute a direct “sign of life” from the individual but arise from administrative or household-level processes, so that individuals may appear in registers despite being absent, leading to false positive observation error. Existing approaches for register data often either rely on ad-hoc decisions that ignore one or both types of observation error, or only offer inference on population snapshots but not on dynamics, or are computationally too slow to be used in practice. We propose a scalable framework for inferring population size and dynamics from register data, building on Cormack-Jolly-Seber type capture-recapture models formulated as hidden Markov models. Inference is carried out using maximum likelihood estimation, with uncertainty quantified via the Bag of Little Bootstraps. The model accounts for temporary emigration, incorporates an arbitrary number of possibly interacting observation registers subject to false positive and false negative observation error, and allows observation probabilities to vary with individual characteristics and unobservable heterogeneity. We illustrate the approach using Swedish population registers, where overcoverage - individuals registered as living in the country although they are no longer present - provides a motivating example. The application yields new insights into population dynamics and individual trajectories, demonstrating the potential of the proposed model for register-based demographic research.
\end{abstract}
 

\section{Introduction}

\subsection{Background}

Reliable inference on population dynamics, encompassing both the monitoring of migration flows and the estimation of total population size, is a cornerstone of effective policymaking, resource allocation, and demographic research \citep{ONS2024FuturePopulationMigration}. This task is particularly challenging for mobile populations, such as migrants, or for cryptic/hard-to-reach populations, such as drug users, for whom traditional censuses are often infeasible, expensive or not timely enough \citep{bohning2018capture, skinner2018issues}. Inaccurate population size estimates, whether for the full population or for specific groups of individuals, lead to biases in key demographic rates, such as birth and death rates \citep{wallace2025long}. Reliable population size estimation is therefore essential for providing policy- and decision-makers with realistic assessments of population needs, allowing effective allocation of resources and planning of interventions. 

As an alternative to a census, an increasing number of countries have adopted a register-based approach for monitoring populations, to either replace or complement censuses \citep{wallgren2014register}. Register data refers to administrative data that is regularly collected and updated via interaction with official bodies, such as the birth, marriage and employment registers. Nordic countries have a long history of a fully register-based system, and in the past $10$ years the official statistics agencies of many other countries have proposed plans to adopt register-based approaches for population size estimation, such as the UK \citep{abbott2020population}, New Zealand \citep{bycroft2015census} and Australia \citep{chipperfield2024robust}. However, any single administrative source is inherently incomplete, as individual registers do not capture every member of the population within a given period (false negative observation error). While the use of multiple registers covering a diverse range of activities improves overall coverage, the probability of non-observation persists across all sources, and hence population size cannot be inferred from the raw data as there exist individuals who do not appear in any register. On the other hand, registers may also contain indirect observations (false positive observation error), defined as records that do not constitute a direct ``sign of life'' from the individual but are instead artefacts of administrative or household-level processes. In such cases, an individual appears on a register despite being absent from the population. Additionally, emigration cannot be assumed permanent, and therefore a period of non-observation followed by subsequent reappearance in registers does not imply continuous residence. The combination of false positive observation error and temporary emigration introduce additional complexity in determining an individual's true location. Importantly, observation probabilities are register- and individual-specific, varying as functions of individual covariates as well as unobserved individual heterogeneity beyond what is explained by observed covariates \citep{wallgren2014register, bohning2018capture, gimenez2018individual, forsythe2021demystifying}. Finally, dependence between registers is often exhibited, such that activity in one register can increase or decrease the probability of appearing in another, as the underlying behaviours are correlated, either positively or negatively, or due to administrative process links. These data challenges motivate the need for tailored statistical approaches for reliable inference from register data.

\subsection{Existing Approaches}

A common, relatively simple but ad-hoc group of approaches to estimate population size and dynamics is based on identifying a ``sign of life'' within administrative registers, alternatively referred to as ``register-trace'' approaches. Sign of life approaches track individuals over a series of administrative registers and only classify individuals that appear on at least one of these registers as present in the country. They have been used in official settings by countries such as Sweden \citep{sweden2015overtackning, sweden2018registration}, Norway \citep{KrokedalNergaardKvalo2024}, Finland \citep{statisticsfinland2024vamuu} and Italy \citep{solari2023statistical}, and proposed as a possible approach by others, such as the UK (specifically England and Wales) \citep{abbott2020population}. Restrictions on the types of registers considered can be made, for example the ``zero personal income approach'' in which individuals with no personal income from a variety of sources in a given year are excluded and assumed to be no longer resident \citep{weitoft1999mortality, aradhya2017repeat}. However, these methods rely heavily on ad-hoc register rules to define population size and cannot monitor population dynamics such as migration, except where movements are officially documented. These ad-hoc methods must be decided in advance and kept consistent, resulting in difficult formalisation and application to other settings. 

A more formal class of methods is based on log-linear models for contingency tables, using the multiple systems estimation (MSE) framework. These models treat the observed register combinations as cells in a contingency table and infer the probability of being observed in each cell. This in turn allows estimation of the number of individuals in the unobserved ``all-zero'' cell, corresponding to the number of individuals who do not appear in any registers, allowing population size estimation. In this setting, recently \cite{mussino2023multiple} allowed for the inclusion of individual covariates and accounted for dependence between registers using interaction terms. In the special case of two administrative registers, dual system estimation (DSE) has been proposed as a way to estimate population size. Recent applications in Australia \citep{chipperfield2024robust} and Ireland \citep{10.1093/jrsssa/qnad065} adapt the DSE framework to the realities of administrative data by addressing two key sources of observation error. First, records that do not correspond to genuine presence in the population are mitigated through trimming which removes implausible or low-quality observations (i.e. false positive observation error). Secondly, linkage problems are handled by incorporating the probability that records belonging to the same individual were correctly linked, given their individual characteristics. MSE approaches are computationally efficient, allow individual- and register-specific observation probabilities to depend on individual covariates, and account for dependence between registers. However, because they operate on an annual basis, they only provide a snapshot of the population size, and cannot be used to infer population dynamics or identify the underlying demographic processes driving these dynamics. A related approach by \cite{yildiz2015models} proposed hierarchical log-linear models with offsets to combine an inaccurate administrative source with an auxiliary data source, but similarly operated on an annual basis and additionally required the existence of an auxiliary data source such as a coverage survey.

Finally, capture-recapture (CR) models provide a natural framework for inferring population dynamics and population size, while accounting for observation errors, and they form the foundation of the approach we develop in this paper. Open‑population CR models such as the Cormack-Jolly-Seber (CJS) framework \citep{cormack1964estimates, jolly1965explicit, seber1965note} condition on an individual’s first capture in the study and model subsequent survival and recapture probabilities, allowing individuals to be followed over time. In classical CR models \citep{pollock, mccrea2014analysis}, individuals are repeatedly ``captured'' and ``recaptured'', generating a capture history that records when each individual is observed. A capture occasion refers to a single opportunity to observe an individual at a given time point. Although CR methods originated in ecology for estimating the size of wildlife populations, they are now widely used in human population studies, including disease prevalence \citep{poorolajal2017using, bohning2020estimating, THOMPSON2023109710}, homelessness estimation \citep{coumans2017estimating}, and other applications in the social and medical sciences.

However, within administrative register settings, the application of CR models has been limited \citep{bohning2018capture}, with the exception of \cite{santos2024using}. The \cite{santos2024using} work demonstrates the potential of CR models to recover latent demographic processes from register data, but also highlights important practical limitations. In particular, their Bayesian framework implementation requires repeatedly sampling individual- and time-specific latent states, and hence is unsuitable for realistic data sizes. It also does not accommodate false positive observation error; as a result, individuals who have emigrated but continue to appear in registers through administrative artefacts are misclassified as present, leading to overestimation of population size. Finally, it represents individual heterogeneity only through observed covariates rather than a flexible latent structure. In practice, however, individuals may differ in their tendency to appear in administrative registers due to behavioural or demographic factors that are not recorded, and such unobserved heterogeneity is a well-known source of biased parameter estimates in CR models \citep{gimenez2010individual}. Taken together, these limitations motivate the need for a new modelling approach tailored to administrative data.

\subsection{Our Contribution} 

We propose a unified framework that is the first to jointly address the key challenges of administrative register data - multiple interacting registers subject to both types of observation error, temporary emigration, and unobserved individual heterogeneity - while remaining scalable to full-population datasets. We employ a hidden Markov model (HMM) formulation \citep{zucchini2009hidden}, which provides a link between an individual's unobserved true states (e.g. present, dead, or emigrated) and their observations, allowing us to efficiently model latent movement dynamics. A key methodological innovation is a multicategory logit model \citep{agresti2007categorical} for the observation probabilities, which removes the standard independence assumption across sampling occasions (registers) and allows observation probabilities to vary with covariates and unobserved heterogeneity via a finite-mixture structure. To ensure scalability for large administrative datasets, we integrate a Bag-of-Little-Bootstraps (BLB) procedure \citep{kleiner2012big, kleiner2014scalable} for uncertainty quantification. Standard bootstrapping approaches easily become computationally infeasible when working with large datasets, as is the case with national administrative data. To our knowledge, BLB has not previously been applied in CR or HMM-based population models, despite its suitability for large-scale likelihood-based inference and ability to handle complex parameter structures that would be nearly impossible in alternative uncertainty quantification methods. Together these contributions provide a novel, unified and flexible modelling framework for register-based inference of population size and dynamics.

\subsection{Motivating case study} 

The motivation for this paper comes from the Swedish population registers, specifically for the migrant population. In Sweden, as well as in many other European countries, all individuals whose actual or planned primary residence is within the country for at least one year are required to register with the Swedish Tax Agency, becoming part of the Register of Total Population (\textit{Registret \"{o}ver totalbefolkningen } - RTB). Upon registration, a personal identification number is assigned, which is necessary for various life activities such as accessing banking and housing; these high incentives to register result in minimal undercoverage (incorrectly excluding individuals from the population) of the \textit{de jure} population, consisting only of individuals awaiting their immigration to be processed. Individuals are equally required to de-register when leaving Sweden, but a combination of lack of knowledge and low incentives mean many individuals do not, resulting in overcoverage (incorrectly retaining individuals in the RTB after their departure from the country) \citep{andersson2023vem}.

In this paper we include a case study where we use Swedish administrative data for all foreign-born adults who first entered the country between $2003 - 2016$. This data is similar to that used by previous papers \citep{mussino2023multiple, santos2024using}, allowing for meaningful comparison of population size estimates across modelling approaches. Our analysis also yields new insights into population trajectories, allowing individual-level movements to be followed over time. While motivated by the Swedish context, the methods proposed are applicable to other settings involving administrative registers. 

This paper is structured as follows. In Section 2 we introduce the proposed modelling approach and inference process. In Section 3 we present the motivating case study and results, including comparison with previous models. In Section 4 we give concluding remarks and discuss some possible extensions for future work. Due to the sensitive and confidential nature of the data used in the case study, the dataset cannot be made publicly available; however, the code has been packaged and openly accessible as the R package \textit{overcoverage} (available at: \url{https://brsantos.github.io/overcoverage/}).

\bigskip
\section{Methods}

\subsection{Observed Data}

We consider longitudinal administrative data on individuals $i = 1, ..., N$ who first enter the study area at some point (year) during the observation period $t = 1, ..., T$. Following the standard CJS approach of conditioning on first entry, we model individuals from the year they first enter the population of interest. The number of new entrants each year is denoted $n_t$. As we condition on first entry, the modelled population at $t = 1$ consists only of those individuals whose entry occurs in that year, so for this cohort we have a population size $N_1 = n_1$. Each year $t$ individuals may appear on a set of administrative registers $k = 1, ..., K$. Therefore, for each individual and year we observe a vector of register indicators $Y_{it} = (Y_{it1}, ..., Y_{itK})$ where $Y_{itk} = 1$ if individual $i$ appears in register $k$ at time $t$ and $0$ otherwise. Each register $k$ within year $t$ constitutes a capture occasion, and the $K$ registers within a given year collectively form the observation record for that year. The observed register pattern $Y_{it}$ does not reveal whether an individual is present, temporarily abroad, permanently emigrated, or deceased. To represent this unobserved and dynamic structure, we introduce a latent state variable and model the data using a hidden Markov model, as in subsequent sections. 

In addition to the $K$ observation registers, we observe administrative event records corresponding to emigration, re-immigration, and death. These registers are treated differently from the $K$ observation registers because they correspond to transitions between latent states, rather than observations conditional on a state, and they are assumed to occur only between years. In particular, an emigration record is (potentially) generated only when an individual transitions from present to abroad, and a death record is (potentially) generated only when an individual transitions into the dead state. Together, these data sources form the input to the HMM described below.

\subsection{Latent State Process}
\label{sec:HMM}
We formulate CR models as HMMs, which marginalise over latent states using the forward algorithm \citep{Laake2013, zucchini2009hidden, jm3}. HMMs are well established in a wide range of fields, with applications in pattern recognition \citep{678085, jm3} and time series modelling, among others \citep{bouguila2022hidden}, including in ecological CR applications \citep{Laake2013, zucchini2009hidden}. In an HMM, the latent state process (e.g. alive, absent, dead), evolves according to a Markov chain, and the observed data is generated from an emission distribution that specifies the probability of each observed outcome given the current latent state.

For each individual $i = 1, ..., N$ and time point $t = 1, ..., T$, let $Z_{it} \in \{1, ..., L\}$ denote the corresponding unobserved true state. Conceptually, the state space denotes whether an individual is (1) alive and present in the study area, (2) alive but abroad, or (3) dead, which is an absorbing state. These broad categories capture the demographic and geographic processes relevant for population estimation.

Individuals initially enter the study alive and present, and subsequently independently evolve according to a Markov process, either remaining present, moving abroad through emigration, returning from abroad through re‑immigration, or dying.

\subsection{Transition Model}

The transition model specifies how individuals move between latent states from one year to the next. The evolution of each individual's latent state between times $t$ and $t+1$ is governed by a set of demographic transition probabilities. These transitions are encoded in an individual- and time-specific $L \times L$ transition matrix $\Gamma_{it}$, whose entries depend on observed covariates. 

We model the probability of three key life-events: survival $s_{it}$, emigration $e_{it}$, and re-immigration $r_{it}$, which are defined such that for a generic life-event, $\theta_{it}$ is the probability that it occurs for individual $i$ transitioning between times $t$ and $t+1$. All three probabilities vary across individuals and time, and are parameterised using logistic regression to incorporate observed sources of heterogeneity \citep{agresti2007categorical}. 

The non-zero entries of $\Gamma_{it}$ are determined by the survival, emigration, and re-immigration probabilities introduced above and the latent state follows a first-order Markov chain 
\begin{equation}
    Z_{i,t+1} | (Z_{i,t} = l) \sim \text{Categorical}(\Gamma_{it}[l, \cdot])
\end{equation}
such that the $l$th row of $\Gamma_{it}$ gives the probabilities of transitioning from state $l$ at time $t$ to each possible state at time $t+1$. This links the individual‑level transition probabilities ($s_{it}$, $e_{it}$ and $r_{it}$) directly to the evolution of the latent state process.

When considering the three conceptual states outlined previously, the transition matrix can be specified as follows:
\begin{equation}
    \Gamma_{it} = \left[ \begin{array}{ccc}
        s_{it}(1-e_{it}) & s_{it}e_{it} & 1-s_{it} \\
        s_{it}r_{it} & s_{it}(1-r_{it}) & 1-s_{it} \\
        0 & 0 & 1
    \end{array}\right] \quad \begin{array}{l}
        \text{(1) Present} \\
        \text{(2) Abroad} \\
        \text{(3) Dead} 
    \end{array}  
    \label{transitionmatrix}
\end{equation}

This structure reflects the demographic processes encoded in the model. Individuals present in the study area may survive and remain present, die, or emigrate, and individuals who are abroad may survive and remain abroad, die abroad, or re-immigrate. The transition matrix is presented here in its most general form but can be tailored to specific case studies depending on the registers that are collected and the administrative events that must be accounted for. 

\subsection{Observation Model}

This section describes how the observation model is constructed conditional on the individual latent state whilst accounting for false negative (Section \ref{sec:f-}) and false positive (Section \ref{sec:f+}) observation error  and observations of migration and death events (Section \ref{sec:emde}).

\subsubsection{Accounting for False Negative Observation Error}
\label{sec:f-}

When an individual is alive and present in the study area the observation $Y_{it}$ corresponds to one of $J$ possible register and covariate combinations, including unobserved in all registers. We incorporate categorical covariates by allowing each register-covariate combination to define its own observable category, effectively extending the emission space in a way analogous to how MSE models handle stratification. We model 
\begin{equation}
    Y_{it} | (Z_{it} = 1) \sim \text{Multinomial}(1; p_{i1t}, ..., p_{iJt})
\end{equation}
where the category probabilities are obtained from a multicategory logit model \citep{agresti2007categorical}. In the baseline-category formulation of multicategory logit models, the log-odds of each category relative to a chosen reference category are modelled as linear functions of covariates. This framework allows the probability of each observation category to depend on individual characteristics, time-varying factors, and interactions between predictors. By treating each register-covariate combination as its own observation category, the model inherits a key strength of MSE approaches: the ability to stratify the observation process across finely defined register cells, allowing dependence between registers, covariate effects, and unobserved heterogeneity to be represented jointly within a single emission model. To our knowledge, this bridging of CR and MSE observation structures has not previously been proposed within a CR or HMM framework.

Let $x_{ijt}$ denote the design vector for observation category $j$ containing indicators for the $K$ registers, covariate categories, and their two-way interactions. The probability of observing category $j$ is
\begin{equation}
     Pr(Y_{it} = j | Z_{it} = 1) = p_{ijt} = \frac{\exp(x_{ijt}^T \gamma)}{\sum_{h=1}^J \exp(x_{iht}^T \gamma)}
\end{equation}

To accommodate unobserved individual heterogeneity in register activity, we introduce a finite mixture model with $G$ latent classes, each with parameter vector $\gamma^{(g)}$ and class‑specific probabilities $p_{ijt}^{(g)}$. Here $\omega_{ig}$ denotes the class membership weight for individual $i$ and class $g$, computed as the probability of belonging to class $g$ conditional on the individual's observed data and current parameter estimations \citep{mcclintock2021worth}. The marginal emission probability is then
\begin{equation}
    p_{ijt} = \sum_{g=1}^G \omega_{ig} p_{ijt}^{(g)}
\end{equation}
capturing differences in register activity patterns that are not explained by observed covariates.

\subsubsection{Accounting for False Positive Observation Error}
\label{sec:f+}

In standard CR models, any recorded observation is typically assumed to imply that the individual was present in the study area at that time. While false negative errors (missed detections) are routinely accommodated, false positive errors (an individual is recorded despite being absent) are rarely modelled. 

Nevertheless, related ideas appear in the literature on misidentification in closed populations, where individuals may be incorrectly recorded as present due to matching errors or ambiguous identifiers \citep{yoshizaki2009modeling, link2010uncovering}. Motivated by this work, and by the fact that administrative registers can be prone to false positive observation error, we extend the CR-HMM framework (open population) to permit individuals who are abroad to appear in some registers with non-zero probability. Specifically, for individuals who are abroad, we define 
\begin{equation}
    Pr(Y_{it} = j | Z_{it} = 2) = q_{ijt}
\end{equation}
where $q_{ijt} = 0$ for register combinations that cannot generate false positive observation errors and $q_{ijt} > 0$ for combinations involving observation only on registers known to produce indirect administrative activity. 

Incorporating these probabilities enables the model to distinguish true presence, temporary emigration, and erroneous administrative traces, which is essential for accurate population size estimation in settings with incomplete de-registration.

\subsubsection{Observations Associated with Migration and Death}
\label{sec:emde}

A death record is generated only when an individual transitions into the dead state. If an individual dies while present, the event is observed with probability $\phi_{it}^p$; if the individual dies while abroad, the event is observed with probability $\phi_{it}^a$. Once the individual has entered the absorbing dead state, no further register activity can occur, and the only possible observation is “no observation”.
\begin{equation} 
    Y_{it} \mid (Z_{it} = 3) = \text{no observation} 
\end{equation}

An emigration event occurs when an individual transitions from the present state to the abroad state, and this event is observed with probability $\psi_{it}^e$. This parameter allows the general model to accommodate settings where emigration may be fully observed, partially observed, or never observed, depending on administrative processes. Once abroad, any appearance in an administrative register would constitute a false positive observation error. A re-immigration event is recorded when an individual transitions from the abroad state to the present state, and this event is observed with probability $\psi_{it}^r$. In the year of re-entry, the individual may also appear on any of the usual $J$ register-covariate combinations, giving rise to an additional set of $J$ observation categories: observed on register-covariate combination $j$ and a re-immigration recorded. 

Taken together, the full observation space consists of $2J + 2$ mutually exclusive categories: $J$ register-covariate combinations when alive and present, one ``emigrated'' category, one ``death registration'' category, and $J$ register-covariate combinations with a re-immigration recorded. 

In practice, the HMM includes additional single-year only intermediate states that ensure migration and death events are recorded in the correct year and that transitions between the conceptual states (present, abroad and dead) are handled in line with the administrative registers. Therefore, the additional ``emigrated'', ``death registration'' and $J$ register-covariate combinations with a re-immigration recorded categories are only observable when an individual is in one of these intermediate states.

\subsection{Inference}

\subsubsection{Parameter Estimation}

For each individual $i$, the likelihood contribution is obtained by marginalising over the latent states using the forward algorithm. Let $\delta_i$ denote the initial state distribution, $P(y_{it})$ the diagonal matrix of emission probabilities for observation $y_{it}$, $\Gamma_{it}$ the transition matrix between times $t$ and $t+1$ and $\mathbf{1}$ a vector of ones. The individual contribution to the likelihood function is
\begin{equation}
    L_i = \delta_i P(y_{i1})\Gamma_{i1}P(y_{i2})...\Gamma_{i,T-1}P(y_{iT})\mathbf{1}
\end{equation}

The full likelihood is the product over individuals, and the log-likelihood is maximised with respect to all model parameters, including the regression coefficients governing the transition probabilities, the mixture-specific observation parameters, and false-positive observation parameters. 

\subsubsection{Uncertainty Quantification}

Although maximum likelihood estimation provides point estimates for all model parameters and the hessian of the log-likelihood yields standard errors for the directly estimated parameters, many quantities of interest are complex non-linear functions of these parameters. Propagating uncertainty to such derived quantities is intractable given the high-dimensional latent state structure, mixture components, and multicategory emission model, making resampling‑based uncertainty quantification necessary.

As administrative register dataset can contain hundreds of thousands of individuals over multiple years, traditional bootstrap methods quickly become computationally infeasible. Bag of Little Bootstraps (BLB) \citep{kleiner2012big, kleiner2014scalable} provides a scalable alternative that preserves statistical accuracy while dramatically reducing computation. Given an original dataset of size $n$, BLB draws $s$ subsets of size $b = n^\gamma$ (with $\gamma \in [0.5, 1]$) without replacement. For each subset, repeated resamples of size $n$ are generated with replacement, but only the $b$ unique individuals in the subset contribute to the likelihood, weighted by their resample counts. Crucially, BLB exploits this fact, allowing the likelihood contribution for each individual to be computed once and then exponentiated according to the number of times that individual appears in the resample. The CR-HMM is fitted to each weighted resample, producing a distribution of parameter estimates for that subset. Aggregating across the $s$ subsets yields point estimates, standard errors and confidence intervals. As each subset can be processed independently, BLB is naturally suited to parallel computation and scales efficiently to full-population administrative data. 

\subsubsection{Latent State Decoding}

Once parameter estimates have been obtained, we reconstruct the most probable latent state sequence for each individual using the Viterbi algorithm \citep{viterbi1967error, forney1973viterbi, zucchini2009hidden, jm3}. This dynamic programming procedure computes, for each individual $i$, time $t$, and state $j$, 
\begin{equation}
    v_{i1}(j) = Pr(z_1 = j) P_j(y_{i1}) \qquad \qquad v_{it}(j) = \max_{s = 1}^S v_{i,t-1}(s) \Gamma_{ijt} P_j(y_{it}) 
    \label{viterbi}
\end{equation}
where $v_{it}(j)$ is the probability of the most likely path ending in state $j$ at time $t$ for individual $i$. For models incorporating finite mixtures in the observation process, the emission probabilities $P_j(y_{it})$ are replaced by mixture-weighted probabilities using the individual-specific weights $\omega_{ig}$.

Backtracking from the most probable final state yields the decoded trajectory $\hat{Z}_{i1}, ..., \hat{Z}_{iT}$ for each individual. These trajectories provide a bootstrap realisation of the annual population size, obtained by counting the number of individuals whose decoded state is present at each time $t$. Repeating this procedure across all BLB resamples produces a distribution of $N_t$ for every year of the study, from which we obtain point estimates, standard errors, and confidence intervals for the population size and any subgroup‑specific quantities of interest.

\section{Case Study} 

\subsection{Data}

We use administrative register data of the Swedish population, including the Register of Total Population (RTB), the Longitudinal Integrated Database for Health Insurance and Labour Market Studies (LISA), the Intergenerational Register, and the Internal and International Moves Register. This data has been collected by different agencies and provided by the Statistics Sweden (SCB), Sweden's official statistics agency. This information is compiled using personal identification numbers and provides insight into various aspects of an individual's life, such as employment status and income, migration history and demographic details. 

Error relating to emigration, and therefore overcoverage, is known to be particularly prevalent among mobile groups of individuals as they are more difficult to ``capture''. While migratory events may not be accurately recorded, life events such as birth and death generally are, resulting in inflation of key demographic rates for specific groups of individuals. In $2001$ Sweden became a member of the Schengen Area and since then immigration has significantly increased, with international migrants making up one fifth of the resident population in $2023$ \citep{mussino2023multiple}. For these reasons we focus on data relating to all foreign-born residents who first entered Sweden as adults between $2003$ and $2016$. In total we work with $721,854$ individuals from $52$ pre-specified countries/country groups such as ``UK and Ireland'', ``India Nepal Bhutan'' and ``North Africa (except Egypt)'', with substantial numbers coming from Eastern Europe, the Middle East and other Nordic countries. For this case study, country of birth is grouped as 1) Denmark and Norway, 2) Eastern Europe, 3) Iceland/Finland, 4) Middle East and North Africa (MENA), 5) United States of America (USA), Canada and Oceania, 6) Western Europe and 7) rest of the World. We also have information about sex, which is treated as binary, as well as age and time since first entering Sweden, which are both treated as categorical. Age is grouped as 1) $18-35$, 2) $36-60$ and 3) over $60$ years old, while time since first entering Sweden is grouped as 1) $0$ (year of entry), 2) $1-5$ years and 3) over $5$ years. These age and country of birth groups have been chosen in line with previous studies for consistency e.g. \citep{mussino2023multiple, santos2024using} allowing comparison, and we have chosen to use these time since first entering Sweden groups following their results regarding migration behaviour.  

Each year an individual is present in the country, we have a record of their observation in ten registers: registration of marriage or registered partnership; registration of divorce or separated partnership; active unemployment, indicating individuals actively searching for work; enrolment in higher education during the autumn term; internal moves within the country; birth of a child; income from pensions, including old-age pension, occupational pension and private pension insurance; employment related earnings, where the sum of earned income and work-related compensation (such as sickness benefit, pregnancy benefit and parental benefit) is greater than zero; social benefits/allowances, including a range of sources such as sickness allowance, parental allowance from the birth or adoption of a child, educational allowance for doctoral students and work disability allowance; and family income, where an individual is part of a household that has income. We also have information regarding an individual's migration history, i.e. immigration, emigration and re-immigration, as well as death records. However, unlike the ten registers outlined above, which capture activity conditional on presence in the country, the migration and death records are subject to the recording limitations described earlier in this section. 

\subsection{Model Specification} 

The general CR-HMM framework described in Section 2 applies directly to the Swedish register data. In this setting, individuals may leave the country either with or without formally de-registering their presence from the country, and the administrative system includes only recorded migration events. A recorded de-registration produces an observed emigration event with certainty, whereas failure to de-register results in overcoverage, where an individual is absent from the country but still appears administratively present. If an individual leaves the country and later re-enters, their re-immigration will be observed with certainty only if they initially de-registered. 

In the general model, absence is represented by a single ``abroad'' state. In the Swedish study, we distinguish four types of absence to reflect the administrative processes governing de-registration and re-registration, and refine the latent state space into the following eight states: (1) present and alive, (2) present and death recorded, (3) abroad and emigration recorded, (4) abroad with known absence, (5) abroad with unknown absence (overcoverage), (6) abroad and death recorded, (7) returned and re-registered, and (8) dead (absorbing). These states allow the model to separate (i) individuals who leave with a recorded emigration (i.e. de-registration), (ii) individuals who leave without de-registering, and (iii) individuals who re-enter the country with a recorded re-registration. In this case study we have two distinct emigration processes that an individual can follow (known vs unknown emigration), and as we are interested in inferring the number of individuals who are abroad without de-registering, we divide the general model's ``abroad'' state to represent these two processes. As in the general model, the additional single-year intermediate states (2, 3, 6, 7) ensure administrative events are recorded in the correct year. 

Transitions between these states are governed by the survival probability $s_{it}$, the emigration probability $e_{it}$, and the re-immigration probability $r_{it}$. In order to distinguish between emigrations that are formally recorded and those that are not, we introduce a de-registration probability $\lambda_{it}$ that denotes the probability that an individual formally de-registers when emigrating. All four probabilities are specified via logistic regression to incorporate individual- and time-varying covariates. Consistent with the annual structure of the registers, we allow at most one migration event per individual per year. The resulting transition matrix is:
\begin{equation}
    \Gamma_{it} = \left[ \begin{array}{cccccccc}
        s_{it}(1-e_{it}) & 1-s_{it} & \lambda_{it}s_{it}e_{it} & 0 & (1-\lambda_{it})s_{it}e_{it} & 0 & 0 & 0 \\
        0 & 0 & 0 & 0 & 0 & 0 & 0 & 1 \\
        0 & 0 & 0 & s_{it}(1-r_{it}) & 0 & 1-s_{it} & s_{it}r_{it} & 0 \\
        0 & 0 & 0 & s_{it}(1-r_{it}) & 0 & 1-s_{it} & s_{it}r_{it} & 0 \\
        s_{it}r_{it} & 0 & 0 & 0 & s_{it}(1-r_{it}) & 1-s_{it} & 0 & 0 \\
        0 & 0 & 0 & 0 & 0 & 0 & 0 & 1 \\
        s_{it}(1-e_{it}) & 1-s_{it} & \lambda_{it}s_{it}e_{it} & 0 & (1-\lambda_{it})s_{it}e_{it} & 0 & 0 & 0 \\
        0 & 0 & 0 & 0 & 0 & 0 & 0 & 1
    \end{array}\right] 
    \label{transitionmatrix2}
\end{equation}

The observation process also follows the general structure described in Section 2. Individuals who are alive and present generate one of $J$ register-covariate combination categories, with probabilities modelled using a multicategory logit model and extended via a finite mixture to capture unobserved heterogeneity. The FMM is incorporated into the ``job income'' register with $G = 2$ latent mixture groups to account for natural differences in individual's propensity to be employed/have income from employment. Migration and death events correspond to additional observation categories (``emigrated'', ``death recorded'', and ``re-immigration \& register pattern''), each associated with the relevant intermediate states. As in the general model, individuals who are abroad may generate false positive observations in registers where administrative activity can occur indirectly, despite the individual's physical absence from the country. In the Swedish case, only the family income register can generate indirect administrative activity: an individual may appear in this register solely because another household member has non-zero personal income. We therefore allow false positive observations only for the category ``unobserved in all registers except family income'', with probabilities $q_{ijt}$ entering the emission matrix for states 5, 6, and 8. 

The general model includes event recording probabilities for emigration, $\psi_{it}^e$, and for re-immigration, $\psi_{it}^r$, corresponding to transitions into the abroad and present states respectively. In the Swedish application, emigration is always recorded when an individual de-registers from the administrative registers (absence is known), and never recorded when an individual fails to de-register (absence is unknown); similarly, an individual's re-immigration/re-registration will only be recorded if they initially de-registered when emigrating. Therefore, we set $\psi_{it}^e = \psi_{it}^r = 1$ when entering/leaving the known abroad state and $\psi_{it}^e = \psi_{it}^r = 0$ when entering/leaving the unknown abroad state. 

Similarly, death registrations are fully observed when present and never observed when abroad (regardless of which abroad state the individual is in), corresponding to $\phi_{it}^p = 1$ and $\phi_{it}^a = 0$. Although, realistically, a small number of individuals die while abroad, the administrative process for registering death outside Sweden is complex and depends on the individual's degree of ongoing attachment to the country. As these cases are rare and require unnecessary additional model complexity, we assume deaths are never registered when absent. 

We fit the model to $721,854$ individuals observed over $14$ years ($2003-2016$), with $10$ administrative registers generating $J = 2^{10}$ mutually exclusive observation categories for each of $18$ possible covariate combinations. The latent process uses the $8$-state structure described above, and the observation model includes $G = 2$ mixture classes to capture unobserved heterogeneity in the ``income from employment'' register only. Parameter estimation proceeds via maximum likelihood using the forward algorithm, with a total of $196$ parameters estimated. Uncertainty is quantified using the Bag of Little Bootstraps. We draw $s = 20$ disjoint partitions of size $b = 36,092$ approximately, and for each subset generate $100$ resamples of size $721,854$, yielding a total of $2,000$ model fits. The model is fit to each of the $20$ partitions in parallel, with each partition taking approximately $5-6$ days to complete. The resulting distribution of parameter estimates provides standard errors and confidence intervals for all transition and observation parameters. Finally, the mixture-weighted Viterbi algorithm is used to decode the most probable latent state sequence for each individual. These trajectories allow us to classify each person as present, known absent, unknown absent (overcoverage), or dead, and to estimate annual population size and subgroup-specific dynamics (e.g. by country of birth, sex, age group, and time since first entering Sweden). The annual population size estimates are then compared against the number of individuals registered in the country each year in the RTB using the following formula to obtain overcoverage estimates. 
\begin{equation}
    \text{OC} = \left(1 - \frac{\text{population size estimate}}{\text{RTB size}} \right) \times 100
\end{equation}

\subsection{Results}

\begin{figure}[!t]
    \centering
    \includegraphics[width=\linewidth]{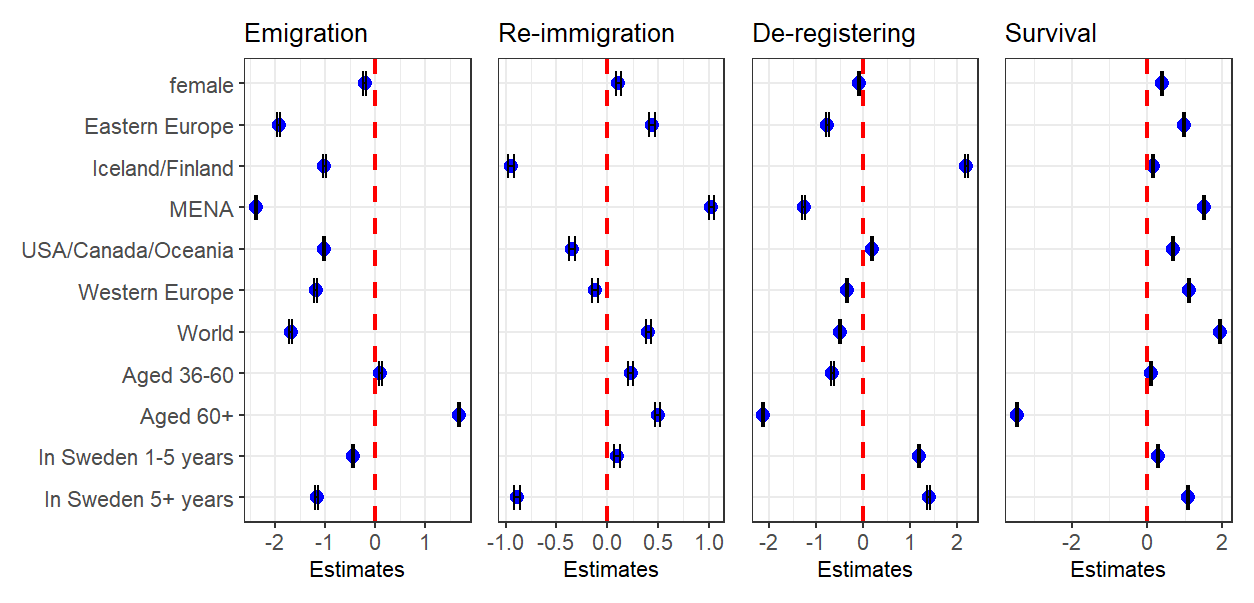}
    \caption{Estimated coefficients for life‑event probabilities with 95\% confidence intervals for each covariate category. Numerical values are provided in the supplementary material.}
    \label{lifeprobabilities}
\end{figure}

Figure \ref{lifeprobabilities} presents the estimated regression coefficients for the four transition probabilities: emigration $e_{it}$, re-immigration $r_{it}$, de-registration $\lambda_{it}$, and survival $s_{it}$, together with $95\%$ confidence intervals. We report effects on the logit scale because the model includes a large number of covariate combinations, and probability-scale plots would compress differences and obscure the relative contribution of each covariate. The logit scale therefore provides a clearer view of the direction and magnitude of covariate effects. 

Across covariates, women exhibit higher survival probabilities than men, consistent with known demographic patterns that women tend to have a longer life expectancy than men. Emigration probabilities vary substantially by country of birth, with individuals from Denmark/Norway (baseline) showing the highest emigration probability, likely reflecting high mobility due to geographical proximity. All other country of birth groups show substantially lower emigration coefficients, with MENA showing the largest negative effect. Individuals from Iceland/Finland show the highest de-registration probability, consistent with the $2004$ Nordic population registration agreement\footnote{https://lovdata.no/dokument/TRAKTAT/traktat/2004-11-01-41}, under which the receiving Nordic country automatically notifies the sending country upon registration, triggering automatic administrative de-registration. By contrast, migrants from Denmark/Norway show lower de-registration rates, which may reflect transnational mobility rather than administrative non-compliance. Individuals who have been in Sweden for $5$ or more years show notably lower emigration and re-immigration probabilities relative to recent arrivals, suggesting increasing settlement over time. Interpretation of all effects must account for population composition, as some covariate groups (e.g. aged $60+$, USA/Canada/Oceania and Iceland/Finland) are small, meaning individual events have a larger influence on estimated probabilities; see the supplementary material for figures relating to the population composition. 

We also estimate the coefficients of the baseline-category logit model governing observation probabilities. Due to the large number of register combinations ($2^{10}$) for each covariate combination, we report the marginal observation probabilities for each register individually rather than for each combination. Table \ref{tab:obs_probs} presents these estimates. As expected, registers corresponding to events that occur only a limited number of times in an individual's life, or only to specific population subgroups (marriage, divorce, pension) have low observation probabilities. The family income register has the highest marginal probability ($0.849$), while pension has the lowest ($0.009$), with the latter explained by the age profile and recent migration history of the study population. When stratified by age, pension observation probabilities follow expected patterns ($0.0009$ for ages $18-35$, $0.0051$ for ages $36-60$ and $0.213$ for ages $60+$).  
\begin{table}[!t]
\caption{Marginal register-level observation probabilities with 95\% confidence intervals.}
\label{tab:obs_probs}
\centering
\begin{tabular}{|c|c|}
\hline
Register & Estimate (CI) \\
\hline\hline
Married        & 0.025 (0.023, 0.026) \\
Divorced       & 0.020 (0.018, 0.021) \\
Active Unemp.  & 0.385 (0.375, 0.398) \\
Studies        & 0.303 (0.292, 0.314) \\
Internal Move  & 0.287 (0.277, 0.295) \\
\hline
\end{tabular}
\quad
\begin{tabular}{|c|c|}
\hline
Register & Estimate (CI) \\
\hline\hline
Child born     & 0.092 (0.087, 0.097) \\
Pension        & 0.009 (0.008, 0.009) \\
Job Income     & 0.513 (0.505, 0.522) \\
Social         & 0.395 (0.382, 0.408) \\
Family Income  & 0.849 (0.842, 0.856) \\
\hline
\end{tabular}
\end{table}

\begin{figure}[!t]
    \centering
    \includegraphics[width=0.85\linewidth]{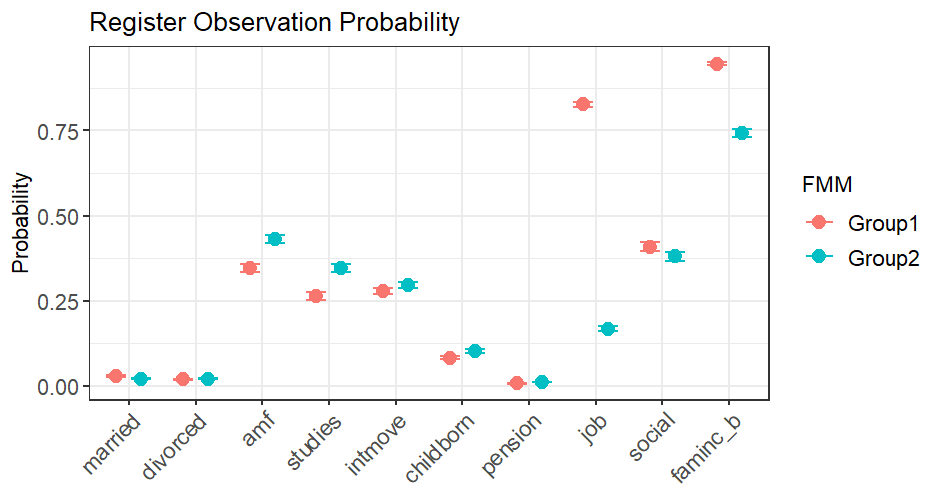}
    \caption{Observation probability estimates for each register, broken down into FMM Group 1 (high job income observation probability) and Group 2 (low job income observation probability), with $95\%$ confidence intervals.}
    \label{fmm_obs}
\end{figure}

The FMM has been incorporated into the observation probabilities for the job income register, capturing unobserved heterogeneity in employment-related activity not explained by observed covariates. We specify $G = 2$ latent mixture classes, estimated as representing groups with high versus low probabilities of being observed on the job income register, with mixing proportion $\hat{\pi} = 0.523$ $(0.517, 0.530)$. It is important to note that the marginal probabilities in Table \ref{tab:obs_probs} are obtained by weighting the class‑specific probabilities (plotted in Figure \ref{fmm_obs}) by the estimated mixing proportion $\hat{\pi}$. Figure \ref{fmm_obs} shows the observation probabilities across all registers broken down by FMM group. The two groups are clearly distinguished on the job income register by construction, but Figure \ref{fmm_obs} reveals that this separation extends meaningfully to other registers. These patterns confirm that the two latent classes represent distinct modes of economic participation rather than a statistical artefact, with Group 1 characterising individuals with sustained labour market attachment and Group 2 characterising those with lower or no employment activity; however, both groups have broadly similar observation probabilities on registers unrelated to employment such as marriage, divorce, internal moves, and birth of a child. 

\begin{figure}[!t]
    \centering
    \includegraphics[width=0.9\linewidth]{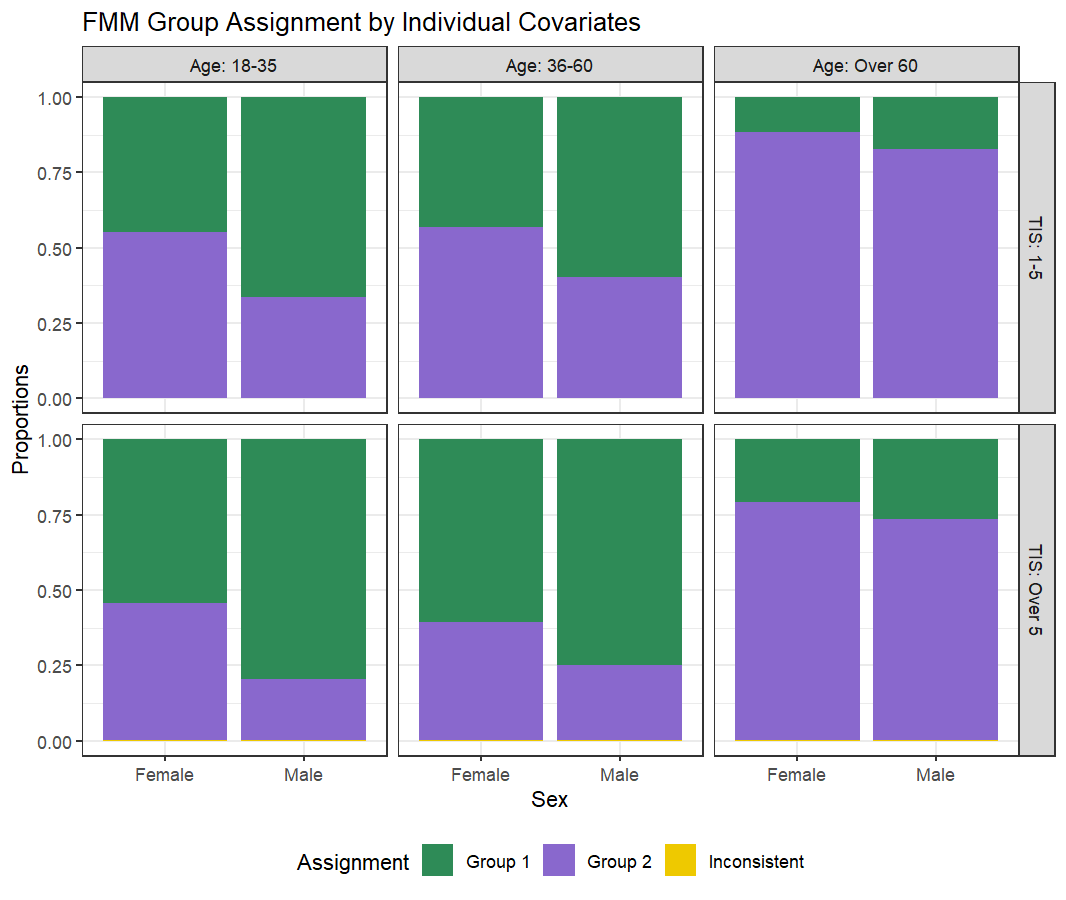}
    \caption{Proportion of registered individuals in $2016$ assigned to FMM groups 1 and 2 consistently ($\geq 90\%$ of bootstraps) and inconsistently, decomposed by sex, age and time since first entering Sweden (TIS).}
    \label{FMM_cov}
\end{figure}

FMM group assignments are highly stable across bootstrap replicates: among individuals registered in $2016$ and conditioning on registration in the country, only $504$ ($<0.1\%$) are inconsistently assigned (defined as $<90\%$ agreement across bootstraps). Figure \ref{FMM_cov} shows the distribution of assignments by sex, age, and time in Sweden (TIS). The pattern is consistent across both TIS groups shown. Men are substantially more likely to belong to the high‑employment‑probability class (Group $1$) than women across all age groups, and this gap is most pronounced for ages $18-35$ and $36-60$. Individuals aged $60+$ are predominantly assigned to Group $2$ regardless of sex, reflecting expected withdrawal from the labour market at older ages. Inconsistent assignments are negligible across all subgroups and are not visible in the figure, confirming the stability of the latent class structure. 

\begin{figure}[!t]
    \centering
    \includegraphics[width=\linewidth]{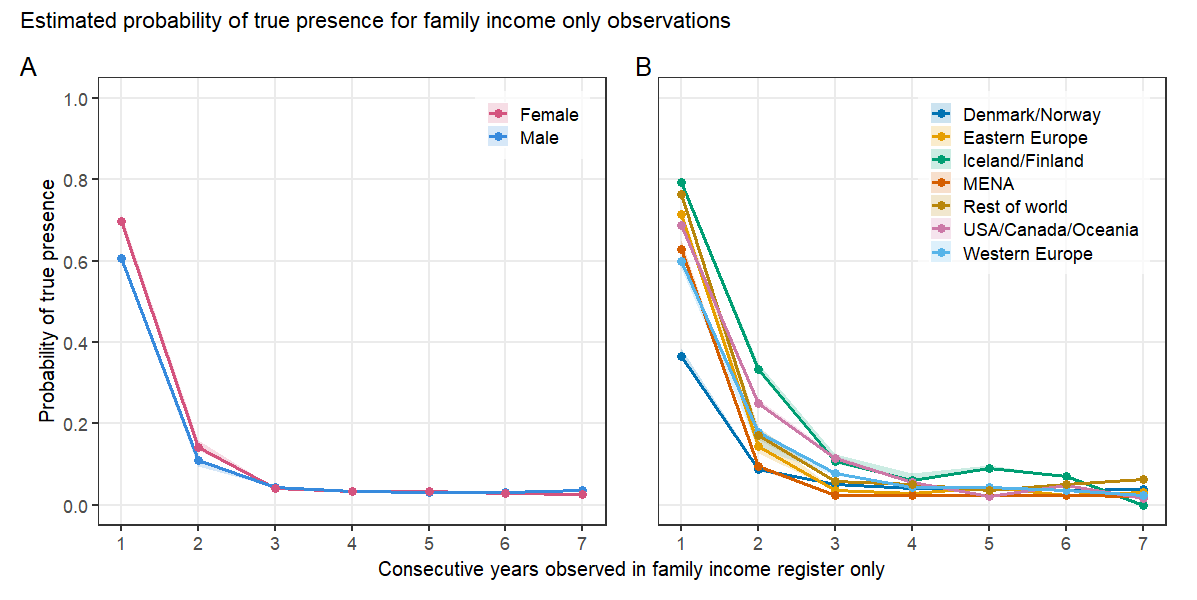}
    \caption{Estimated probability of true presence for individuals observed only in the family income register, as a function of consecutive years of such observations. Panel A shows results by sex and panel B by country of birth. Shaded bands show $95\%$ confidence intervals.}
    \label{fp_consec}
\end{figure}

We examine individuals who appear only in the family income register in at least one year (the register for which false positive observation error is modelled) and estimate the probability that each such observation corresponds to true presence versus absence from the country. Figure \ref{fp_consec} presents these estimated probabilities as a function of the number of consecutive years of family-income-only observation, decomposed by sex (panel A) and country of birth (panel B). The count of individual-year observations underlying each estimate is provided in the supplementary material and cells with small counts of observations should be interpreted with caution. The most striking feature of both panels is the sharp decline in the probability of true presence between years 1 and 2, with probabilities remaining low and approximately stable from year 3 onwards. A single year of family-income-only observation is therefore much more likely to reflect genuine presence than two or more consecutive years, which are increasingly indicative of physical absence combined with continued familial ties to Sweden. Women show a higher probability of true presence than men at year 1 ($0.697$ vs $0.606$), but both groups converge to near zero by year 3. This is consistent with the disproportionate representation of women among family-income-only observations noted earlier, where financial dependence on a household member's income is more plausibly associated with co-residence than with absence. Substantial variation by country of birth is also observed. Denmark/Norway stands out as the group with the lowest probability of true presence at year 1 ($0.365$), consistent with the transnational mobility patterns discussed in the transition probability results. At the other extreme, Iceland/Finland shows the highest probability of true presence at year 1 ($0.794$), consistent with the Nordic registration agreement but crucially these individuals do not benefit geographically in the same way individuals from Denmark/Norway do - since individuals from Iceland/Finland would very likely be automatically de-registered upon emigrating, the model correctly infers that their continued appearance in any register is more plausibly explained by genuine presence. 

\begin{figure}[!t]
    \centering
    \includegraphics[width=0.9\linewidth]{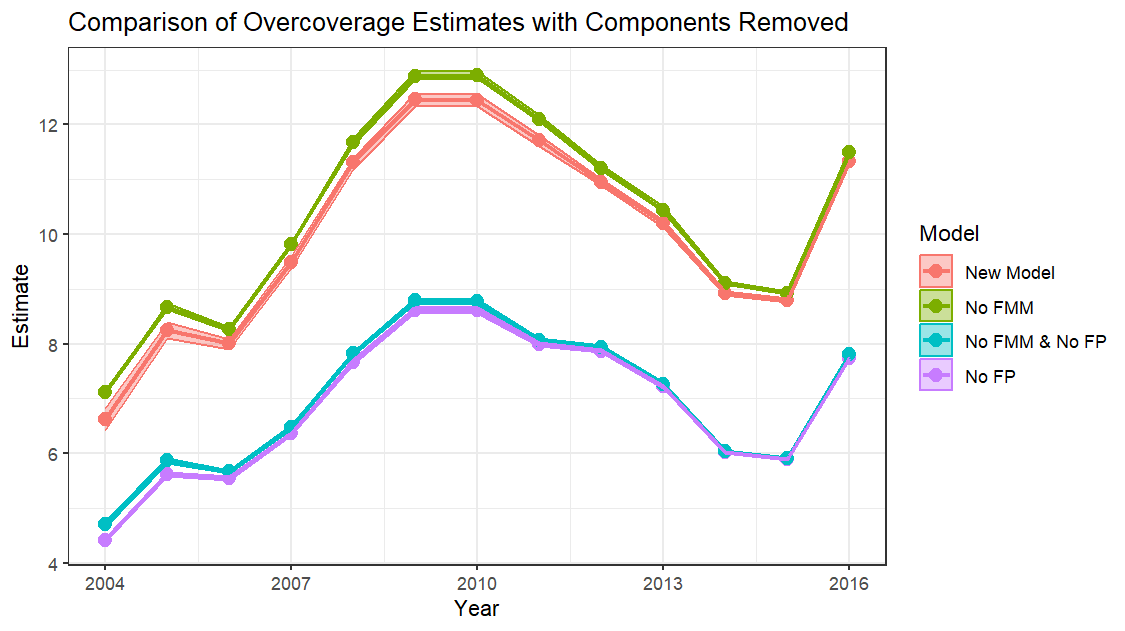}
    \caption{Overcoverage estimates over time for the full proposed model and three reduced variants obtained by removing the FMM, false positive observation error modelling, or both. Shaded bands show $95\%$ confidence intervals.}
    \label{OC_comparison}
\end{figure}

To assess the contribution of the FMM and false positive modelling components, we re‑fitted the model to a $5\%$ subsample (to reduce computational time) under three reduced specifications: (i) without the FMM, (ii) without false‑positive observation error modelling, and (iii) without both. Figure \ref{OC_comparison} compares overcoverage estimates over time across these variants. All four models show a broadly consistent temporal pattern, with overcoverage rising to a peak in $2009-2010$, declining through $2015$, and rising again in $2016$. However, the level of overcoverage differs substantially across model variants. The two models that omit false positive observation error (No FP and No FMM \& No FP) produce estimates approximately $3-4\%$ lower than the full model throughout the study period, demonstrating that allowing individuals to appear in the family‑income register while physically absent is a major driver of increased overcoverage. The model without the FMM but retaining false positive modelling (No FMM) produces estimates very close to the full model, with only a small upward shift, indicating that the FMM has a more modest but still consistent effect on overcoverage. The component comparison therefore confirms that both additions are contributing genuine signal rather than noise.

Taken together, the results demonstrate that the proposed model recovers a richer and more nuanced picture of population dynamics than is possible with MSE or sign-of-life approaches. The transition probability estimates are broadly consistent with those of \cite{santos2024using}, providing reassurance that the two models are recovering the same underlying demographic patterns despite differences in the observation process, inference framework, and sample size. Overcoverage estimates follow the same temporal pattern, with both approaches identifying a peak around $2009-2010$ followed by a decline; however, our estimates are systematically higher throughout the study period, likely due to the explicit modelling of false positive observation errors as illustrated in Figure \ref{OC_comparison}.

Additional results provide further insight into the activity of specific (groups of) individuals in the country and the prevalence of specific events. Direct comparison for the FMM groups and covariate groups for each register can be found in the supplementary material, alongside plots relating to the assignment of individuals to each FMM group and false positive observation state. The $95\%$ confidence intervals for all model parameter estimates are very narrow due to the large sample size; the supplementary material contains tables equivalent to all figures presented in this section.

\section{Discussion}

We have developed an efficient, scalable and highly flexible framework for population size estimation and inferring population dynamics using incomplete, overlapping administrative registers. By formulating a CJS-type CR model with a HMM structure, fitted using BLB, we enable efficient marginalisation over latent states and allow the model to be fitted to full-population administrative data. The framework accommodates temporary emigration, multiple interacting registers via a multicategory logit observation model, false positive and false negative observation errors, and individual heterogeneity in the observation process.

Several methodological contributions distinguish this work from existing approaches. Unlike \cite{santos2024using}, who relied on a Bayesian MCMC approach requiring sampling of all latent states and therefore analysed only a $5\%$ sample, our HMM formulation combined with the BLB enables full-population inference. The computational gains arise from marginalising over latent states via the forward algorithm and exploiting parallel computation, making the approach feasible for national-scale administrative data. Our multicategory observation model captures dependence between multiple registers within a year, extending standard CR assumptions and aligning with MSE-type modelling of list interactions. Crucially, by embedding this observation structure within an open-population CR framework, the model delivers what neither MSE nor sign-of-life approaches can: longitudinal individual trajectories, the ability to track migration dynamics over time, and the ability to distinguish genuine presence from administrative artefacts. Sign-of-life approaches rely on ad-hoc register rules and cannot identify individuals who remain administratively present after departure; MSE approaches operate annually and cannot recover the underlying demographic processes driving population change. The present framework addresses both limitations within a unified model.

Our results highlight the richness of the individual-level inference produced by the model. We find substantial variation in migration dynamics across demographic groups, for example, the model yields predominantly age-specific survival estimates as a direct output of the framework, while country of birth is strongly associated with different propensities for emigration, re-immigration, and de-registration, potentially reflecting differing migration motives. The observation model further reveals how individuals interact with administrative systems, with finite-mixture components capturing subgroups who differ in their probability of appearing in specific registers (in the case study this is the income from employment register). This extends previous findings \citep{mussino2023multiple, santos2024using} by providing individual-specific allocations to latent states and mixture components, as well as identifying individuals likely to generate false positive observations. 

Our overcoverage estimates are higher than those reported in previous work \citep{mussino2023multiple, santos2024using}, and this warrants discussion. We believe this reflects, at least in part, the explicit modelling of false positive observation errors -- by allowing individuals who are abroad to generate administrative activity through indirect register processes such as family income, the model avoids misclassifying administrative artefacts as genuine presence. Models that do not account for false positive observation errors will tend to retain individuals in the present state for longer, underestimating overcoverage. Comparing model variants in Figure \ref{OC_comparison} supports this interpretation. In particular, the false positive observation model has a larger effect on overcoverage estimates than the finite mixture component, suggesting that practitioners in similar settings should prioritise modelling indirect register activity when register quality is imperfect. Nevertheless, our estimates remain notably lower than those produced by deterministic register-trace/sign-of-life approaches, with and without the family income register, as shown in the supplementary material. Care must be taken to ensure that the equivalent version of the presented model is used for this comparison (i.e. without and with false positive observation error modelling respectively). 

Several assumptions and modelling choices merit further discussion. The model assumes at most one migration event per individual per year; for highly mobile subgroups, such as individuals from neighbouring Nordic countries engaging in short-term cross-border employment, this assumption may be violated. However, as most administrative registers are recorded annually, a finer temporal resolution is not available within the current data structure, and the annual timescale represents a practical constraint rather than a modelling choice. On identifiability, the model involves a complex latent state space, finite mixture components, false positive observation parameters, and a large observation space. Identifiability is supported in our setting by the richness of the register data (ten registers observed annually over fourteen years, with substantial individual-level covariate information), but in settings with fewer registers or shorter observation windows some parameters may be weakly identified, and sensitivity analysis would be advisable. Heterogeneity is currently included in one register (income from employment); extending this to multiple registers or to life-event transitions may improve realism but could introduce identifiability or computational challenges. Register interactions are limited to two-way terms, consistent with standard MSE practice; higher-order interactions could in principle be included but would raise similar concerns. 

Our estimates of population size and overcoverage depend on the set of registers included and on the treatment of false positive observation errors. The registers used here are ``active'' registers requiring individuals to engage in specific activities to be observed. Certain subpopulations such as home-makers or retirees with foreign pensions, may therefore remain undetected. Extending the register set to include more passive lists, such as hospital or police records, could reduce this uncertainty and improve coverage of otherwise unobserved individuals. Additionally, while we follow previous work in focusing on the migrant population who enter between $2003-2016$, the framework could be extended with additional data, resulting in estimates more representative of the full migrant population. Similarly, this work could be extended to the full population of Sweden by also considering Swedish born residents. 

Our use of a CJS-type model conditions on the first observed entry into the population. For the Swedish case study, undercoverage at entry (i.e. the individual enters the country unobserved) is believed to be negligible. In settings where individuals may enter unobserved, the framework could be extended to a Jolly-Seber formulation which infers entry times \citep{jolly1965explicit, seber1965note}. We restrict attention to the period $2003-2016$ because earlier data is of lower quality and list definitions change over time. Starting in $2003$ ensures consistent data and avoids inflating the sample with long-term residents who are more detectable and whose true entry times are unknown. 

The generalisability of this framework beyond Sweden deserves consideration. The Swedish application benefits from several features that make it particularly well-suited to this approach: a comprehensive system of linked administrative registers, universal personal identification numbers enabling reliable record linkage, and strong institutional coverage across the population. We are currently applying the model to equivalent administrative data from Norway, which will provide a direct test of generalisability across two countries with similar, well-established register-based systems and will allow comparison of migration dynamics under consistent modelling assumptions. Transferability to countries with weaker register infrastructure, less reliable linkage, or fewer available registers may require simplification of the observation model or additional sensitivity analysis, and we would recommend beginning with a reduced specification in such settings before introducing the full model complexity. 

Although individual-level inference is computationally intensive, parallel computation is increasingly accessible, however, any model will eventually encounter computational limits as resources are finite and much larger population data is common. A natural next step is to develop a population-level MSE framework that incorporates the key components of our CR-HMM model. We are currently working on this extension, with the aim of providing a more streamlined and scalable alternative for official statistics settings, while retaining the capacity to model complex longitudinal dependence structures and observation errors. 

Overall, this work demonstrates that detailed, high-resolution modelling of administrative registers is both feasible and informative at national scale. By combining HMM marginalisation, BLB, and an MSE-type observation structure, we provide a powerful and scalable approach for understanding population dynamics in register-based systems, opening new avenues for methodological developments and practical application.

\subsection*{Acknowledgments}
The computations and data handling were enabled by resources provided by the National Academic Infrastructure for Supercomputing in Sweden (NAISS), partially funded by the Swedish Research Council through grant agreement no. 2022-06725.

\subsection*{Funding}
This research was supported by the Swedish Research Council (VR), grant number 2021-00875. 
The first author was supported by a doctoral scholarship from the Migration and Movement Signature Research Theme, University of Kent.
The third author was partially financed by national funds through FCT - Fundação para a Ciência e a Tecnologia under the projects UID/00006/2025 and UID/PRR/00006/2025.

\subsection*{Supplementary Material}
\textbf{Supplement to ``A capture-recapture hidden Markov model framework for register-based inference of population size and dynamics''}

The supplementary material provides additional supporting material for the main paper. It includes figures providing further detail on model specification, numerical tables corresponding to all figures presented in the paper, and additional model results.

\bibliographystyle{plainnat}
\bibliography{bibliography}

\end{document}


\maketitle 

\subsection*{Hidden Markov Model}

\begin{figure}[H]
    \begin{center}
    \begin{tikzpicture}[
        roundnode/.style={circle, draw=black!60, very thick, minimum size=1cm}
    ]
        \node[roundnode] (state1) {$z_1$};
        \node[roundnode] (state2) [right=of state1] {$z_2$};
        \node[roundnode] (state3) [right=of state2] {$z_3$};
        \node[roundnode] (state4) [right=of state3] {$z_T$};
 
        \node[roundnode] (obs1) [above=of state1] {$y_1$};
        \node[roundnode] (obs2) [above=of state2] {$y_2$};
        \node[roundnode] (obs3) [above=of state3] {$y_3$};
        \node[roundnode] (obs4) [above=of state4] {$y_T$};
 
        \draw[->] (state1.north) -- (obs1.south);
        \draw[->] (state2.north) -- (obs2.south);
        \draw[->] (state3.north) -- (obs3.south);
        \draw[->] (state4.north) -- (obs4.south);
 
        \draw[->] (state1.east) -- (state2.west);
        \draw[->] (state2.east) -- (state3.west);
        \draw[dashed, ->] (state3.east) -- (state4.west);
 
        \draw node at (-2,0) {State};
        \draw node at (-2,2) {Observation};
        \draw node at (0,3)  {$t = 1$};
        \draw node at (2,3)  {$t = 2$};
        \draw node at (4,3)  {$t = 3$};
        \draw node at (5,3)  {$\dots$};
        \draw node at (6,3)  {$t = T$};
    \end{tikzpicture}
    \end{center}
    \caption{Dependence graph for the observation and latent state time series for HMMs. States depend only on the previous state, while each observation is fully dependent on the current state.}
    \label{hmm}
\end{figure}
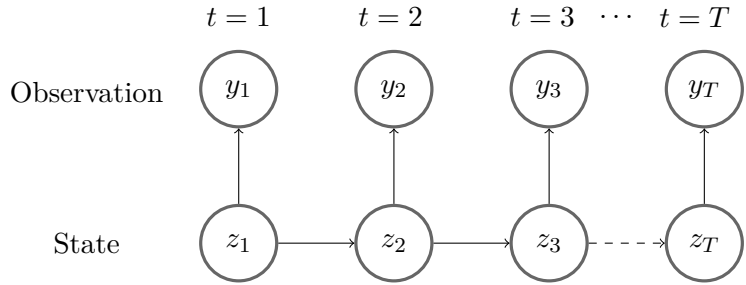

\subsection*{State and Observation Process}

\begin{figure}[H]
    \centering
    \includegraphics[width=0.9\linewidth]{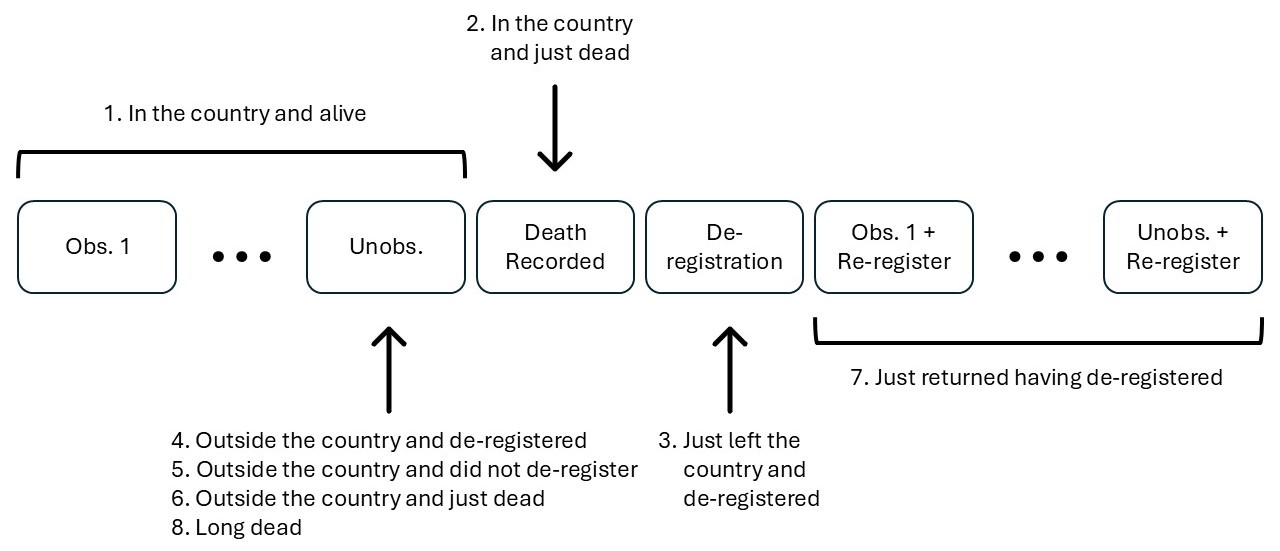}
    \caption{This diagram illustrates the possible observations when considering the Swedish register data. These $2^{(R+1)}+2$ observations specify the columns of the observation matrix $\Omega$ (where $R$ is the number of registers considered), indicating which of the $8$ states can produce each available observation.}
    \label{obs}
\end{figure}

\begin{figure}[H]
    \centering
    \includegraphics[width=1.35\linewidth, angle=270]{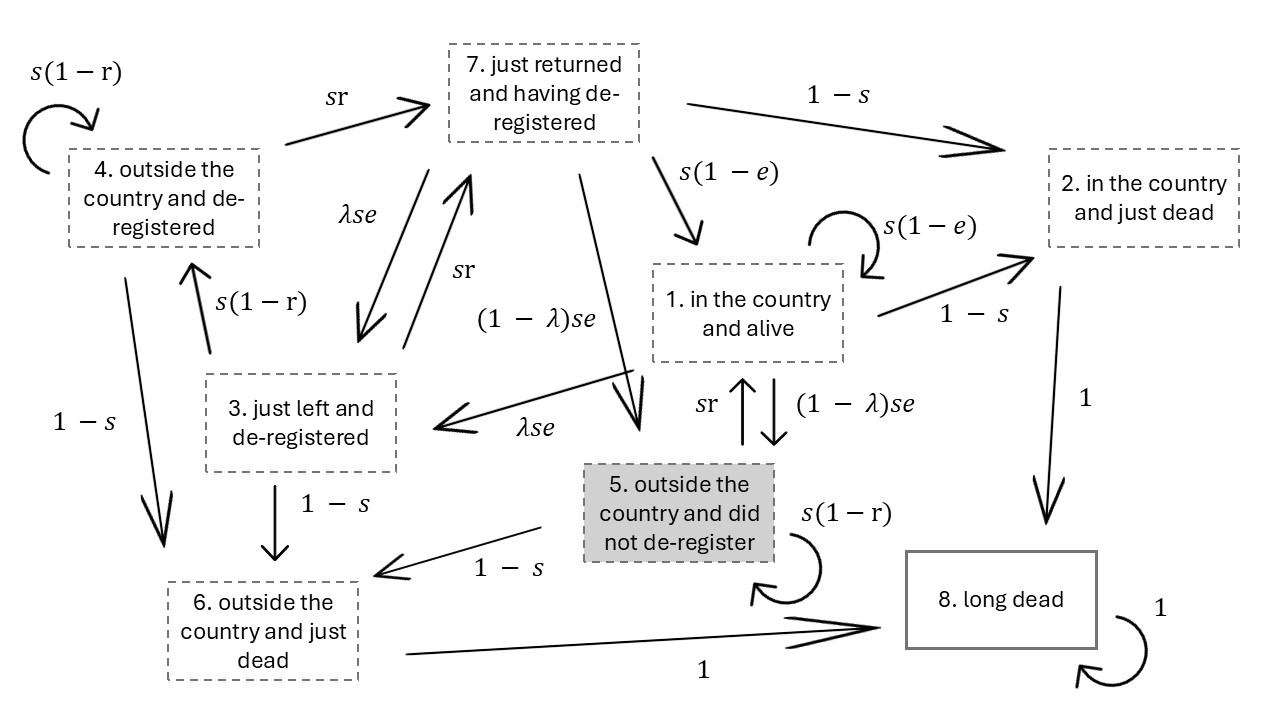}
    \caption{This diagram illustrates how individuals can transition between each of the $8$ states we consider in our case study, along with the corresponding parameters. The parameters used are defined as follows: $e$ is the emigration probability, $r$ is the re-immigration probability, $\lambda$ is the probability of de-registering when leaving the country, and $s$ is the survival probability. The state ``long dead'' is absorbing, highlighted by the solid box boundary, and the state ``outside the country and did not de-register'', filled in grey, corresponds to over-covered individuals.}
    \label{states}
\end{figure}
 
\subsection*{Forward Algorithm} 

For each individual $i$ and time point $t$, the HMM is defined by an initial state distribution $\delta_i = (\delta_{i1}, ..., \delta_{iS})$, a transition matrix $\Gamma_{it}$ of dimension $S \times S$ where $\Gamma_{it}(a,b) = Pr(Z_{i,t+1} = b | Z_{i,t} = a)$, and an emission distribution $P(y_{it})$ which is a diagonal matrix such that $[P(y_{it})]_{jj} = Pr(Y_{it} = y_{it} | Z_{it} = j)$. The likelihood contribution for individual $i$ is obtained by marginalising over all possible latent state sequences. Using the forward algorithm, we define 

\begin{equation}
    \alpha_{i1} = \delta_i P(y_{i1}) \qquad \qquad \alpha_{it} = \alpha_{i,t-1} \Gamma_{i,t-1} P(y_{it})
\end{equation}

and the marginal likelihood is 

\begin{equation}
    L_i = \alpha_{iT} \textbf{1}
\end{equation}

This formulation allows efficient computation even with large state spaces and individual-specific transition matrices, and forms the basis for the transition and observation models.

\subsection*{Bag of Little Bootstraps}

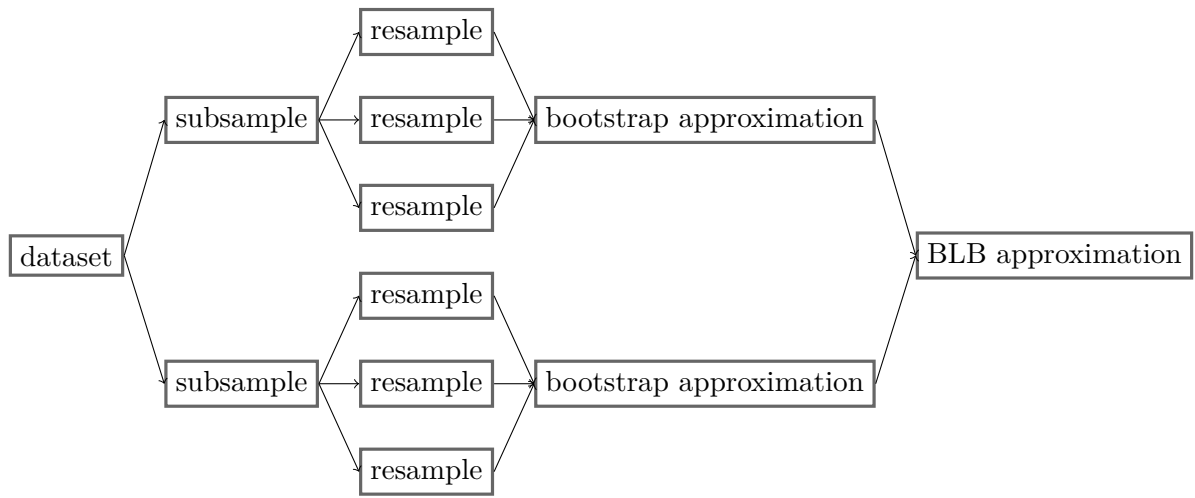
\begin{figure}[H]
    \begin{center}
    \begin{tikzpicture}[
textnode/.style={rectangle, draw=black!60, very thick, node distance=15pt}
]
        \node[textnode] (resample1) {resample};
        \node[textnode] (resample2) [below=of resample1] {resample};
        \node[textnode] (resample3) [below=of resample2] {resample};
        \node[textnode] (resample4) [below=of resample3] {resample};
        \node[textnode] (resample5) [below=of resample4] {resample};
        \node[textnode] (resample6) [below=of resample5] {resample};

        \node[textnode] (subsample1) [left=of resample2] {subsample};
        \node[textnode] (subsample2) [left=of resample5] {subsample};

        \node[textnode] (dataset) [left=of subsample2, yshift=1.7cm] {dataset};

        \node[textnode] (approximation1) [right=of resample2] {bootstrap approximation};
        \node[textnode] (approximation2) [right=of resample5] {bootstrap approximation};

        \node[textnode] (BLB) [right=of approximation2, yshift=1.7cm] {BLB approximation};

        \draw[->] (subsample1.east) -- (resample1.west);
        \draw[->] (subsample1.east) -- (resample2.west);
        \draw[->] (subsample1.east) -- (resample3.west);
        \draw[->] (subsample2.east) -- (resample4.west);
        \draw[->] (subsample2.east) -- (resample5.west);
        \draw[->] (subsample2.east) -- (resample6.west);
        \draw[->] (dataset.east) -- (subsample1.west);
        \draw[->] (dataset.east) -- (subsample2.west);
        \draw[->] (resample1.east) -- (approximation1.west);
        \draw[->] (resample2.east) -- (approximation1.west);
        \draw[->] (resample3.east) -- (approximation1.west);
        \draw[->] (resample4.east) -- (approximation2.west);
        \draw[->] (resample5.east) -- (approximation2.west);
        \draw[->] (resample6.east) -- (approximation2.west);
        \draw[->] (approximation1.east) -- (BLB.west);
        \draw[->] (approximation2.east) -- (BLB.west);
        
    \end{tikzpicture}
    \end{center}
    \caption{A simplified flowchart showing the steps required for the Bag of Little Bootstraps (BLB) approximation. Generalisation to a larger number of subsamples, for example $10$, and a reasonable number of resamples, for example $100$, is necessary for implementation.} 
\end{figure}

\section*{Demographics}
\begin{figure}[H]
    \centering
    \includegraphics[width=0.8\linewidth]{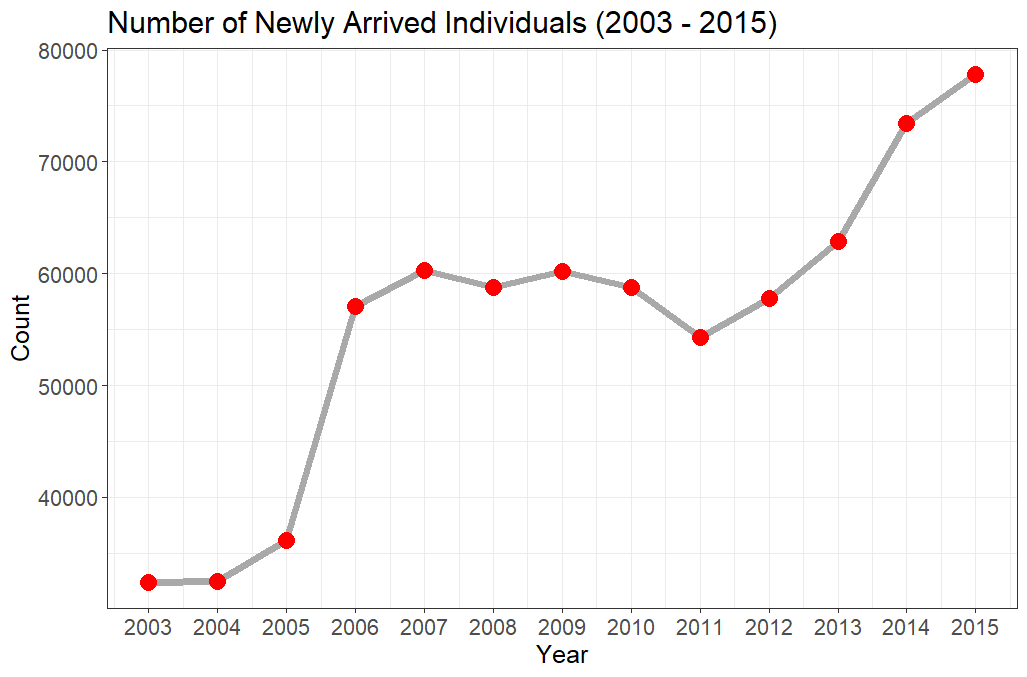}
    \caption{The number of newly arrived individuals in each year of the study $2003 - 2015$ (individuals are observed in $2016$ but there are no new arrived that year). In the first year of the study ($2003$), only individuals who enter that year are part of our population of interest, and thus the true population size in $2003$ is known.} 
    \label{demographics_new}
\end{figure}

\begin{figure}[H]
    \centering
    \includegraphics[width=0.9\linewidth]{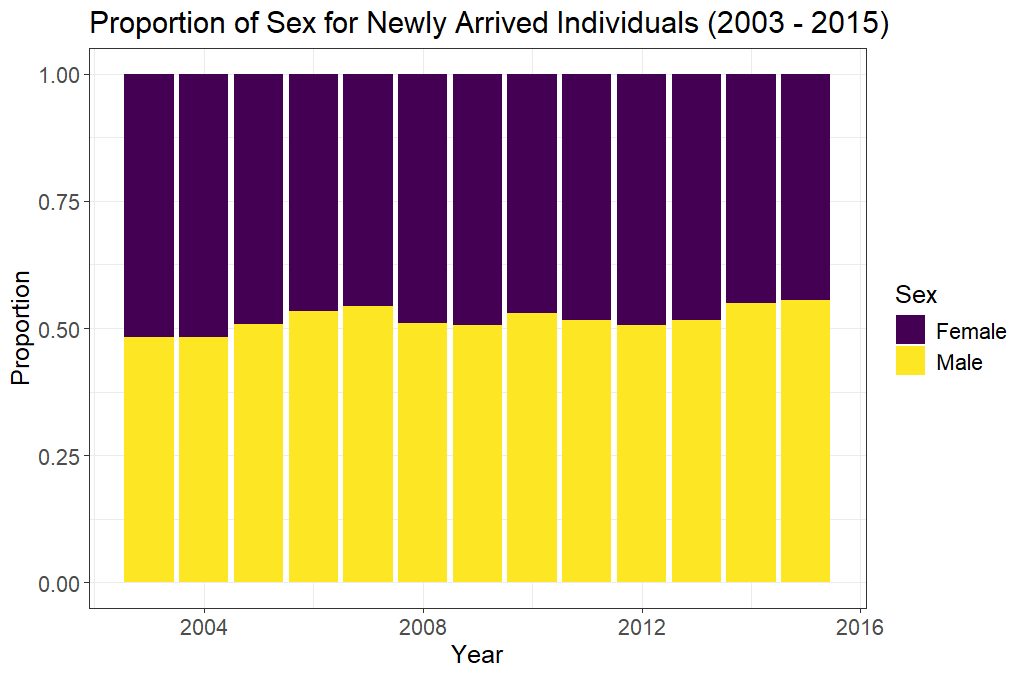}
    \caption{The distribution of sex for newly arrived individuals in each year of the study $2003 - 2015$ (individuals are observed in $2016$ but there are no new arrived that year).} 
    \label{demographics_sex}
\end{figure}

\begin{figure}[H]
    \centering
    \includegraphics[width=0.9\linewidth]{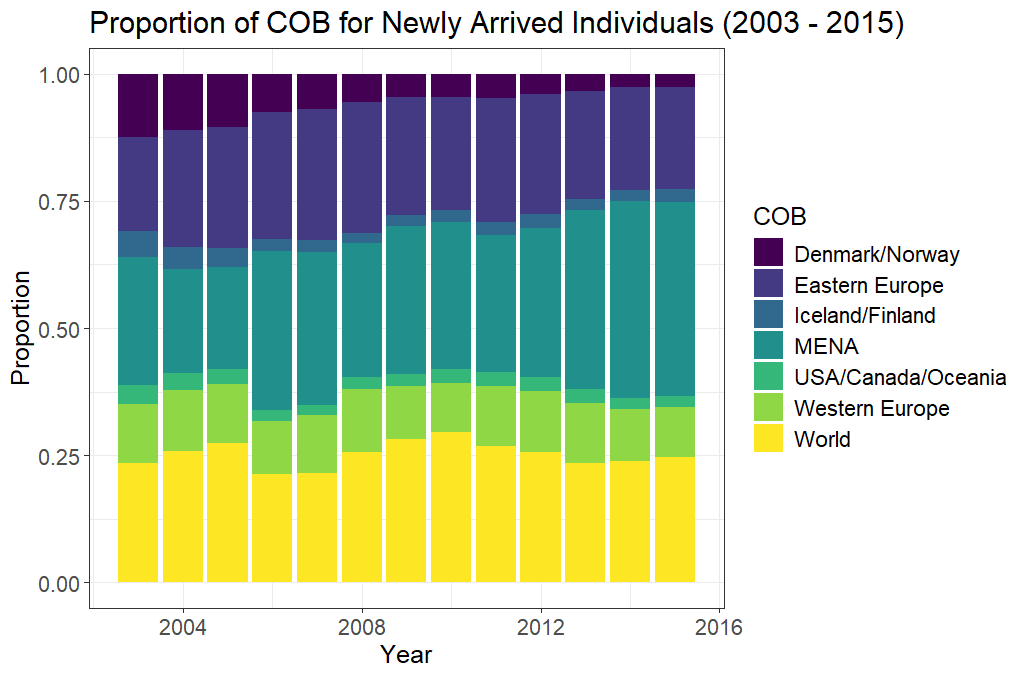}
    \caption{The distribution of country of birth for newly arrived individuals in each year of the study $2003 - 2015$ (individuals are observed in $2016$ but there are no new arrived that year).} 
    \label{demographics_cob}
\end{figure}

\begin{figure}[H]
    \centering
    \includegraphics[width=0.85\linewidth]{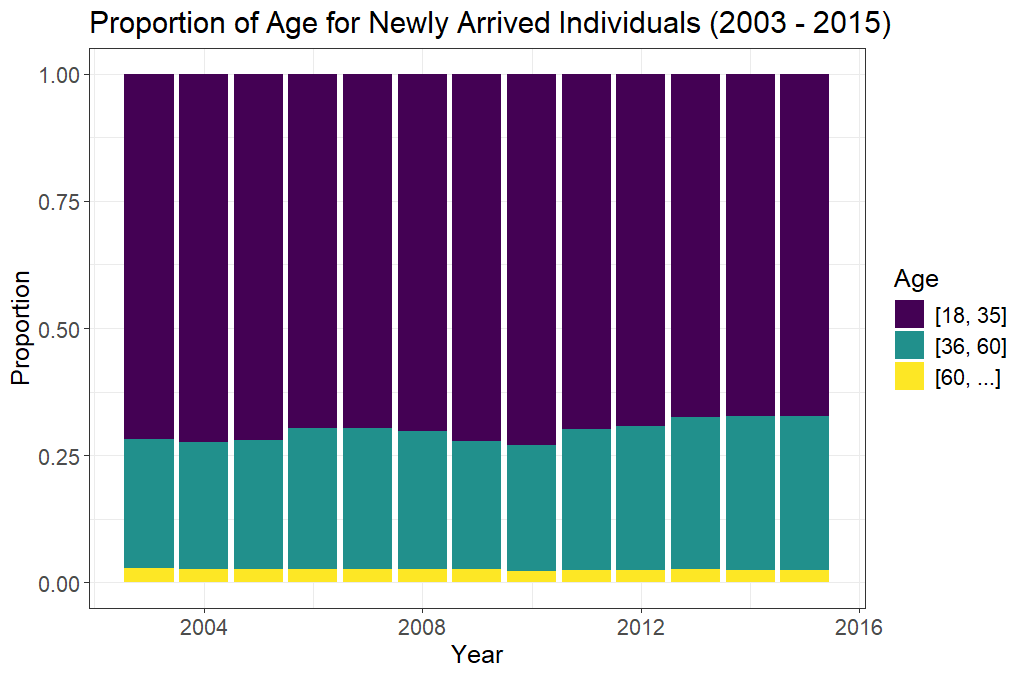}
    \caption{The distribution of age for newly arrived individuals in each year of the study $2003 - 2015$ (individuals are observed in $2016$ but there are no new arrived that year).} 
    \label{demographics_age}
\end{figure}

The following tables show the number of individuals observed on each register being considered, for each year of the study $2003-2016$. The final line of the table show the counts of individuals observed on only the family income register each year - these individuals motivate the incorporation of uncertain sightings. 

\begin{table}[H]
\caption{Counts of individuals observed in each register, 2003--2016.}
\label{tab:register_counts_0309}
\centering
\small
\begin{tabular}{@{} l rrrrrrr @{}}
\toprule
Register & 2003 & 2004 & 2005 & 2006 & 2007 & 2008 & 2009 \\
\midrule
Married        &     0 &  1,084 &  1,792 &  2,800 &  4,696 &  6,211 &  7,404 \\
Divorced       &     0 &    404 &  1,013 &  2,079 &  3,425 &  4,487 &  5,703 \\
Active Unemp.  & 5,853 & 15,028 & 23,905 & 40,309 & 55,528 & 67,895 & 86,978 \\
Studies        & 5,319 & 13,103 & 21,333 & 33,412 & 38,270 & 45,891 & 57,405 \\
Internal Move  & 8,263 & 17,448 & 26,151 & 44,562 & 58,013 & 69,237 & 80,106 \\
Child born     & 2,176 &  5,757 &  8,721 & 12,570 & 16,864 & 20,328 & 24,440 \\
Pension        &     4 &     27 &     58 &    132 &    284 &    487 &    799 \\
Job Income     & 8,366 & 20,662 & 35,647 & 57,982 & 86,138 & 112,877 & 130,730 \\
Social         & 8,366 & 20,662 & 35,647 & 57,982 & 86,138 & 112,877 & 130,730 \\
Family Income  & 21,333 & 44,306 & 68,622 & 109,579 & 151,852 & 190,631 & 225,511 \\
\addlinespace
Only FamInc    & 5,129 &  7,261 &  9,731 & 14,013 & 19,809 & 24,291 & 27,668 \\
\bottomrule
\end{tabular}
\vspace{1em}
\begin{tabular}{@{} l rrrrrrr @{}}
\toprule
Register & 2010 & 2011 & 2012 & 2013 & 2014 & 2015 & 2016 \\
\midrule
Married        &  8,030 &  8,520 &  9,827 & 10,626 & 11,293 & 11,562 & 11,765 \\
Divorced       &  7,137 &  7,588 &  8,362 &  9,298 &  9,612 &  9,566 &  9,535 \\
Active Unemp.  & 111,248 & 127,193 & 146,330 & 170,500 & 194,234 & 218,563 & 209,938 \\
Studies        & 67,494 & 70,608 & 130,128 & 150,115 & 168,826 & 182,401 & 162,430 \\
Internal Move  & 88,512 & 94,881 & 102,660 & 113,899 & 132,424 & 148,615 & 133,284 \\
Child born     & 28,406 & 30,052 & 32,979 & 36,126 & 40,151 & 42,790 & 41,802 \\
Pension        &  1,211 &  1,870 &  2,611 &  3,482 &  4,639 &  6,204 &  9,757 \\
Job Income     & 157,488 & 192,282 & 220,979 & 249,754 & 284,603 & 327,152 & 354,098 \\
Social         & 157,488 & 192,282 & 220,979 & 249,754 & 284,603 & 327,152 & 354,098 \\
Family Income  & 263,320 & 302,518 & 340,768 & 382,755 & 434,683 & 491,390 & 496,451 \\
\addlinespace
Only FamInc    & 27,860 & 29,728 & 24,161 & 26,121 & 28,708 & 31,122 & 28,702 \\
\bottomrule
\end{tabular}
\end{table}

\newpage
\section*{Additional Results}

\subsection*{Life Event Probabilities}

Estimated coefficients for life event probabilities with their $95\%$ confidence intervals for each covariate category. These estimates are plotted on the logit scale and the baseline category has estimates $e: 0.481$ $(0.475, 0.487)$, $r: 0.084$ $(0.082, 0.086)$, $\lambda: 0.529$ $(0.522, 0.536)$ and $s: 0.996$ $(0.995, 0.996)$.

\begin{table}[H]
\caption{Estimated covariate effects for the emigration $e$, re-immigration $r$, de-registration $\lambda$ and survival $s$ probabilities, with 95\% confidence intervals.}
\centering
\begin{tabular}{@{} l cc @{}}
\toprule
Covariate & Emigration & Re-immigration \\
\midrule
Female              & $-0.211$ ($-0.240$, $-0.185$) & $\phantom{-}0.109$ ($0.084$, $0.136$) \\
Eastern Europe      & $-1.932$ ($-1.960$, $-1.904$) & $\phantom{-}0.442$ ($0.416$, $0.473$) \\
Iceland/Finland     & $-1.018$ ($-1.044$, $-0.990$) & $-0.947$ ($-0.974$, $-0.918$) \\
MENA                & $-2.385$ ($-2.413$, $-2.356$) & $\phantom{-}1.026$ ($1.003$, $1.052$) \\
USA/Canada/Oceania  & $-1.027$ ($-1.052$, $-0.999$) & $-0.349$ ($-0.376$, $-0.319$) \\
Western Europe      & $-1.195$ ($-1.220$, $-1.167$) & $-0.122$ ($-0.149$, $-0.094$) \\
World               & $-1.687$ ($-1.715$, $-1.659$) & $\phantom{-}0.407$ ($0.381$, $0.436$) \\
Aged 36--60         & $\phantom{-}0.102$ ($0.075$, $0.129$)  & $\phantom{-}0.233$ ($0.203$, $0.259$) \\
Aged 60+            & $\phantom{-}1.671$ ($1.642$, $1.696$)  & $\phantom{-}0.497$ ($0.469$, $0.524$) \\
In Sweden 1--5 yrs  & $-0.446$ ($-0.470$, $-0.419$) & $\phantom{-}0.101$ ($0.072$, $0.131$) \\
In Sweden 5+ yrs    & $-1.174$ ($-1.201$, $-1.146$) & $-0.890$ ($-0.920$, $-0.861$) \\
\bottomrule
\end{tabular}
\vspace{1em}
\begin{tabular}{@{} l cc @{}}
\toprule
Covariate & De-registration & Survival \\
\midrule
Female              & $-0.093$ ($-0.117$, $-0.064$) & $\phantom{-}0.394$ ($0.367$, $0.420$) \\
Eastern Europe      & $-0.764$ ($-0.794$, $-0.738$) & $\phantom{-}0.984$ ($0.957$, $1.011$) \\
Iceland/Finland     & $\phantom{-}2.188$ ($2.163$, $2.216$)  & $\phantom{-}0.163$ ($0.137$, $0.191$) \\
MENA                & $-1.270$ ($-1.297$, $-1.240$) & $\phantom{-}1.508$ ($1.483$, $1.535$) \\
USA/Canada/Oceania  & $\phantom{-}0.183$ ($0.156$, $0.213$)  & $\phantom{-}0.683$ ($0.655$, $0.711$) \\
Western Europe      & $-0.349$ ($-0.376$, $-0.322$) & $\phantom{-}1.111$ ($1.085$, $1.140$) \\
World               & $-0.493$ ($-0.523$, $-0.464$) & $\phantom{-}1.949$ ($1.922$, $1.976$) \\
Aged 36--60         & $-0.661$ ($-0.687$, $-0.633$) & $\phantom{-}0.101$ ($0.072$, $0.130$) \\
Aged 60+            & $-2.126$ ($-2.155$, $-2.101$) & $-3.460$ ($-3.489$, $-3.431$) \\
In Sweden 1--5 yrs  & $\phantom{-}1.178$ ($1.151$, $1.205$)  & $\phantom{-}0.301$ ($0.274$, $0.329$) \\
In Sweden 5+ yrs    & $\phantom{-}1.386$ ($1.357$, $1.416$)  & $\phantom{-}1.087$ ($1.062$, $1.113$) \\
\bottomrule
\end{tabular}
\end{table}

\subsection*{Marginal Observation Probabilities}

The multicategory logit emission model is defined over all possible register–covariate combinations. Let $X$ denote the design matrix enumerating these combinations, and let $p_j$ denote the estimated emission probability for category $j$ (obtained from the fitted multicategory logit model).

For register $k$, the marginal probability of observing an individual in that register is obtained by summing over all joint categories in which register $k$ is observed:
$$Pr(Y = k | Z = \text{Present}) = \sum_{j: X_{j,k} = 1} p_j$$
\begin{table}[H]
\caption{Marginal register-level observation probabilities for the two finite mixture groups, alongside their $95\%$ confidence intervals.}
\centering
\begin{tabular}{@{} l cc @{}}
\toprule
Register & Group 1: Mean (95\% CI) & Group 2: Mean (95\% CI) \\
\midrule
Married        & 0.028 (0.027, 0.030) & 0.021 (0.020, 0.022) \\
Divorced       & 0.019 (0.018, 0.020) & 0.020 (0.019, 0.021) \\
AMF            & 0.344 (0.334, 0.357) & 0.430 (0.419, 0.443) \\
Studies        & 0.264 (0.253, 0.274) & 0.346 (0.335, 0.358) \\
Internal move  & 0.279 (0.269, 0.288) & 0.295 (0.286, 0.304) \\
Child born     & 0.082 (0.077, 0.086) & 0.103 (0.097, 0.108) \\
Pension        & 0.007 (0.006, 0.007) & 0.011 (0.011, 0.012) \\
Job income     & 0.827 (0.819, 0.834) & 0.168 (0.161, 0.176) \\
Social         & 0.408 (0.395, 0.421) & 0.380 (0.367, 0.394) \\
Family income  & 0.946 (0.943, 0.950) & 0.743 (0.732, 0.753) \\
\bottomrule
\end{tabular}
\end{table}

We also obtain estimates for the probability of being unobserved: for Group 1: $0.023$ $(0.022, 0.025)$, for Group 2: $0.111$ $(0.105, 0.116)$, and overall: $0.065$ $(0.062, 0.068)$.

For covariate specific categories (sex, age group, time in Sweden), we compute conditional probabilities by normalising within the relevant subset of categories. For example, for sex:
$$Pr(Y = k | Z = \text{Present}, \text{Male}) = \frac{\sum_{j: X_{j,k} = 1, X_{j,\text{sex}} = 0} p_j}{\sum_{j: X_{j,\text{sex}} = 0} p_j}$$
Analogous expressions are used for age and time‑in‑Sweden groups.
\begin{figure}[H]
  \centering
  \includegraphics[width=0.8\textwidth]{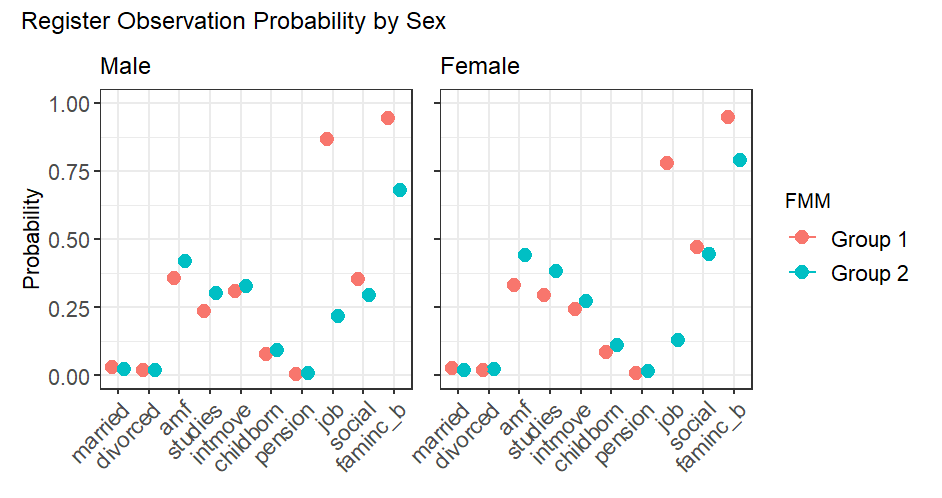}
  \caption{Observation probabilities for each register for each sex covariate group (male and female), broken down into FMM groups in which FMM Group 1 is more likely to be observed in the register relating to income from a job, while FMM Group 2 is less likely.}
  \label{obs_sex_bysex}
\end{figure}
\begin{figure}[H]
  \centering
  \includegraphics[width=\textwidth]{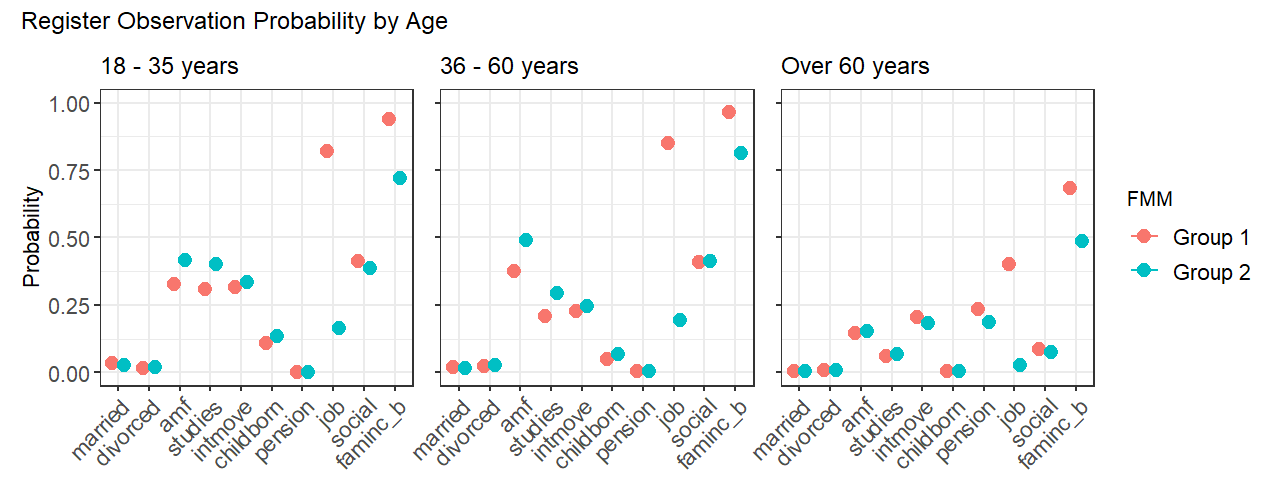}
  \caption{Observation probabilities for each register for each age covariate group ($18-35$, $36-60$ and over $60$), broken down into FMM groups in which FMM Group 1 is more likely to be observed in the register relating to income from a job, while FMM Group 2 is less likely.}
  \label{obs_age_byage}
\end{figure}
\begin{figure}[H]
  \centering
  \includegraphics[width=\textwidth]{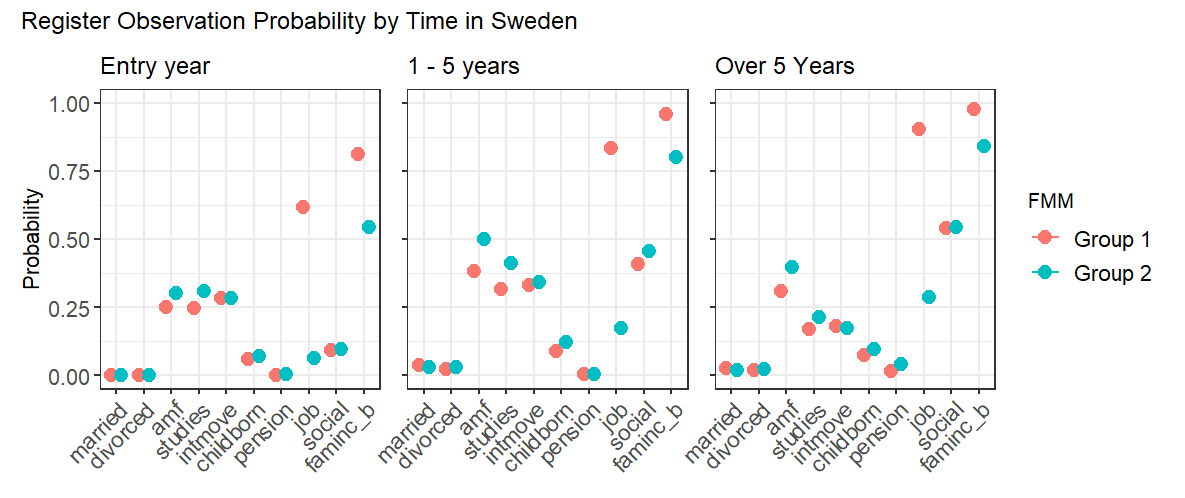}
  \caption{Observation probabilities for each register for each time in Sweden covariate group (entry year, $1-5$ years and over $5$ years), broken down into FMM groups in which FMM Group 1 is more likely to be observed in the register relating to income from a job, while FMM Group 2 is less likely.}
  \label{obs_tis_bytis}
\end{figure}

\begin{table}[H]
    \centering
    \caption{Overall and finite mixture group specific conditional register-level observation probabilities for the sex covariate, alongside their $95\%$ confidence intervals.}
    \subcaption*{\textbf{Male}}
    \begin{tabular}{@{} l ccc @{}}
        \toprule
        Register & Overall & Group 1 & Group 2 \\
        \midrule
        Married       & 0.026 (0.025, 0.027) & 0.030 (0.029, 0.032) & 0.021 (0.020, 0.022) \\
        Divorced      & 0.018 (0.017, 0.019) & 0.018 (0.017, 0.019) & 0.018 (0.017, 0.019) \\
        AMF           & 0.385 (0.375, 0.397) & 0.355 (0.345, 0.368) & 0.418 (0.407, 0.430) \\
        Studies       & 0.268 (0.258, 0.277) & 0.237 (0.227, 0.247) & 0.301 (0.292, 0.311) \\
        Internal move & 0.317 (0.308, 0.326) & 0.310 (0.299, 0.319) & 0.326 (0.317, 0.334) \\
        Child born    & 0.085 (0.081, 0.090) & 0.078 (0.074, 0.083) & 0.093 (0.089, 0.098) \\
        Pension       & 0.007 (0.007, 0.008) & 0.006 (0.005, 0.006) & 0.009 (0.008, 0.009) \\
        Job income    & 0.558 (0.550, 0.567) & 0.869 (0.862, 0.874) & 0.218 (0.209, 0.228) \\
        Social        & 0.326 (0.315, 0.338) & 0.354 (0.341, 0.366) & 0.295 (0.284, 0.307) \\
        Family income & 0.819 (0.812, 0.826) & 0.946 (0.943, 0.949) & 0.679 (0.669, 0.691) \\
        \bottomrule
    \end{tabular}
    \vspace{1em}
    \subcaption*{\textbf{Female}}
    \begin{tabular}{@{} l ccc @{}}
        \toprule
        Register & Overall & Group 1 & Group 2 \\
        \midrule
        Married       & 0.023 (0.022, 0.024) & 0.026 (0.024, 0.027) & 0.020 (0.019, 0.021) \\
        Divorced      & 0.021 (0.019, 0.022) & 0.020 (0.019, 0.021) & 0.021 (0.020, 0.023) \\
        AMF           & 0.383 (0.371, 0.397) & 0.332 (0.320, 0.346) & 0.440 (0.427, 0.455) \\
        Studies       & 0.336 (0.323, 0.349) & 0.295 (0.282, 0.308) & 0.381 (0.367, 0.394) \\
        Internal move & 0.257 (0.246, 0.266) & 0.243 (0.232, 0.252) & 0.272 (0.261, 0.282) \\
        Child born    & 0.097 (0.092, 0.103) & 0.086 (0.080, 0.091) & 0.110 (0.104, 0.117) \\
        Pension       & 0.010 (0.010, 0.011) & 0.008 (0.007, 0.008) & 0.013 (0.012, 0.014) \\
        Job income    & 0.469 (0.460, 0.478) & 0.778 (0.768, 0.789) & 0.129 (0.123, 0.136) \\
        Social        & 0.459 (0.445, 0.474) & 0.471 (0.457, 0.487) & 0.446 (0.431, 0.462) \\
        Family income & 0.873 (0.866, 0.880) & 0.947 (0.943, 0.951) & 0.792 (0.781, 0.802) \\
        \bottomrule
    \end{tabular}
\end{table}
 
\begin{table}[H]
    \centering
    \caption{Overall and finite mixture group specific conditional register-level observation probabilities for the age covariate, alongside their $95\%$ confidence intervals.}
    \subcaption*{\textbf{Aged 18--35}}
    \begin{tabular}{@{} l ccc @{}}
        \toprule
        Register & Overall & Group 1 & Group 2 \\
        \midrule
        Married       & 0.030 (0.029, 0.032) & 0.034 (0.033, 0.036) & 0.025 (0.024, 0.026) \\
        Divorced      & 0.017 (0.016, 0.018) & 0.017 (0.016, 0.018) & 0.018 (0.017, 0.019) \\
        AMF           & 0.370 (0.359, 0.382) & 0.327 (0.316, 0.340) & 0.417 (0.405, 0.429) \\
        Studies       & 0.352 (0.340, 0.363) & 0.307 (0.295, 0.318) & 0.400 (0.388, 0.412) \\
        Internal move & 0.326 (0.316, 0.335) & 0.317 (0.307, 0.327) & 0.335 (0.325, 0.344) \\
        Child born    & 0.119 (0.113, 0.125) & 0.107 (0.101, 0.113) & 0.133 (0.126, 0.140) \\
        Pension       & 0.001 (0.001, 0.001) & 0.001 (0.001, 0.001) & 0.001 (0.001, 0.001) \\
        Job income    & 0.508 (0.500, 0.517) & 0.822 (0.815, 0.830) & 0.164 (0.157, 0.171) \\
        Social        & 0.400 (0.388, 0.414) & 0.414 (0.401, 0.427) & 0.385 (0.372, 0.399) \\
        Family income & 0.836 (0.829, 0.843) & 0.941 (0.937, 0.944) & 0.721 (0.710, 0.732) \\
        \bottomrule
    \end{tabular}
    \vspace{1em}
    \subcaption*{\textbf{Aged 36--60}}
    \begin{tabular}{@{} l ccc @{}}
        \toprule
        Register & Overall & Group 1 & Group 2 \\
        \midrule
        Married       & 0.018 (0.017, 0.019) & 0.020 (0.019, 0.021) & 0.015 (0.014, 0.016) \\
        Divorced      & 0.024 (0.022, 0.025) & 0.023 (0.021, 0.024) & 0.025 (0.024, 0.027) \\
        AMF           & 0.431 (0.419, 0.445) & 0.377 (0.365, 0.391) & 0.490 (0.477, 0.504) \\
        Studies       & 0.251 (0.240, 0.262) & 0.210 (0.200, 0.220) & 0.296 (0.284, 0.308) \\
        Internal move & 0.236 (0.226, 0.245) & 0.227 (0.217, 0.236) & 0.245 (0.235, 0.254) \\
        Child born    & 0.057 (0.054, 0.061) & 0.048 (0.045, 0.051) & 0.067 (0.063, 0.071) \\
        Pension       & 0.005 (0.005, 0.005) & 0.005 (0.005, 0.006) & 0.005 (0.005, 0.005) \\
        Job income    & 0.538 (0.529, 0.547) & 0.851 (0.843, 0.858) & 0.194 (0.185, 0.204) \\
        Social        & 0.412 (0.399, 0.427) & 0.411 (0.397, 0.426) & 0.414 (0.400, 0.429) \\
        Family income & 0.893 (0.887, 0.899) & 0.965 (0.963, 0.968) & 0.813 (0.804, 0.823) \\
        \bottomrule
    \end{tabular}
    \vspace{1em}
    \subcaption*{\textbf{Aged over 60}}
    \begin{tabular}{@{} l ccc @{}}
        \toprule
        Register & Overall & Group 1 & Group 2 \\
        \midrule
        Married       & 0.004 (0.004, 0.004) & 0.005 (0.005, 0.005) & 0.003 (0.003, 0.003) \\
        Divorced      & 0.007 (0.006, 0.007) & 0.007 (0.007, 0.008) & 0.006 (0.006, 0.007) \\
        AMF           & 0.150 (0.143, 0.157) & 0.146 (0.140, 0.153) & 0.154 (0.147, 0.162) \\
        Studies       & 0.063 (0.060, 0.067) & 0.060 (0.057, 0.064) & 0.067 (0.063, 0.070) \\
        Internal move & 0.194 (0.187, 0.202) & 0.206 (0.197, 0.214) & 0.182 (0.175, 0.190) \\
        Child born    & 0.003 (0.003, 0.003) & 0.003 (0.003, 0.003) & 0.003 (0.003, 0.003) \\
        Pension       & 0.213 (0.202, 0.223) & 0.235 (0.224, 0.247) & 0.188 (0.178, 0.198) \\
        Job income    & 0.222 (0.213, 0.231) & 0.400 (0.384, 0.414) & 0.027 (0.026, 0.029) \\
        Social        & 0.081 (0.077, 0.086) & 0.087 (0.083, 0.092) & 0.075 (0.070, 0.080) \\
        Family income & 0.589 (0.576, 0.604) & 0.683 (0.669, 0.697) & 0.486 (0.473, 0.502) \\
        \bottomrule
    \end{tabular}
\end{table}
 
\begin{table}[H]
    \centering
    \caption{Overall and finite mixture group specific conditional register-level observation probabilities for the time since first entering Sweden covariate, alongside their $95\%$ confidence intervals.}
    \subcaption*{\textbf{Entry year in Sweden}}
    \begin{tabular}{@{} l ccc @{}}
        \toprule
        Register & Overall & Group 1 & Group 2 \\
        \midrule
        Married       & 0.001 (0.001, 0.001) & 0.001 (0.001, 0.002) & 0.036 (0.034, 0.038) \\
        Divorced      & 0.001 (0.001, 0.001) & 0.001 (0.001, 0.001) & 0.023 (0.021, 0.024) \\
        AMF           & 0.276 (0.268, 0.284) & 0.251 (0.244, 0.260) & 0.384 (0.372, 0.399) \\
        Studies       & 0.276 (0.268, 0.285) & 0.246 (0.238, 0.255) & 0.318 (0.304, 0.330) \\
        Internal move & 0.284 (0.276, 0.291) & 0.283 (0.275, 0.291) & 0.331 (0.319, 0.341) \\
        Child born    & 0.065 (0.062, 0.068) & 0.060 (0.057, 0.063) & 0.090 (0.084, 0.096) \\
        Pension       & 0.002 (0.002, 0.003) & 0.002 (0.002, 0.002) & 0.002 (0.002, 0.003) \\
        Job income    & 0.353 (0.345, 0.361) & 0.616 (0.605, 0.628) & 0.834 (0.826, 0.841) \\
        Social        & 0.093 (0.088, 0.097) & 0.091 (0.087, 0.095) & 0.410 (0.396, 0.425) \\
        Family income & 0.684 (0.675, 0.693) & 0.813 (0.805, 0.820) & 0.960 (0.957, 0.963) \\
        \bottomrule
    \end{tabular}
    \vspace{1em}
    \subcaption*{\textbf{1--5 years in Sweden}}
    \begin{tabular}{@{} l ccc @{}}
        \toprule
        Register & Overall & Group 1 & Group 2 \\
        \midrule
        Married       & 0.033 (0.031, 0.035) & 0.001 (0.001, 0.001) & 0.030 (0.028, 0.031) \\
        Divorced      & 0.025 (0.024, 0.027) & 0.001 (0.001, 0.001) & 0.028 (0.026, 0.030) \\
        AMF           & 0.440 (0.428, 0.455) & 0.303 (0.294, 0.311) & 0.501 (0.489, 0.516) \\
        Studies       & 0.362 (0.348, 0.374) & 0.309 (0.301, 0.318) & 0.410 (0.396, 0.423) \\
        Internal move & 0.337 (0.325, 0.347) & 0.285 (0.277, 0.292) & 0.343 (0.332, 0.353) \\
        Child born    & 0.104 (0.098, 0.111) & 0.070 (0.067, 0.073) & 0.120 (0.113, 0.128) \\
        Pension       & 0.004 (0.004, 0.004) & 0.003 (0.003, 0.003) & 0.006 (0.005, 0.006) \\
        Job income    & 0.519 (0.511, 0.528) & 0.063 (0.061, 0.066) & 0.175 (0.167, 0.183) \\
        Social        & 0.431 (0.418, 0.445) & 0.095 (0.090, 0.100) & 0.454 (0.440, 0.469) \\
        Family income & 0.884 (0.878, 0.890) & 0.543 (0.532, 0.553) & 0.800 (0.791, 0.810) \\
        \bottomrule
    \end{tabular}
    \vspace{1em}
    \subcaption*{\textbf{Over 5 years in Sweden}}
    \begin{tabular}{@{} l ccc @{}}
        \toprule
        Register & Overall & Group 1 & Group 2 \\
        \midrule
        Married       & 0.023 (0.022, 0.024) & 0.025 (0.024, 0.027) & 0.020 (0.019, 0.021) \\
        Divorced      & 0.021 (0.020, 0.023) & 0.020 (0.019, 0.021) & 0.023 (0.021, 0.024) \\
        AMF           & 0.350 (0.337, 0.363) & 0.309 (0.297, 0.321) & 0.395 (0.381, 0.409) \\
        Studies       & 0.190 (0.181, 0.200) & 0.170 (0.161, 0.178) & 0.213 (0.203, 0.223) \\
        Internal move & 0.176 (0.169, 0.184) & 0.179 (0.171, 0.187) & 0.174 (0.166, 0.181) \\
        Child born    & 0.085 (0.080, 0.090) & 0.075 (0.071, 0.079) & 0.096 (0.091, 0.101) \\
        Pension       & 0.027 (0.025, 0.029) & 0.016 (0.015, 0.017) & 0.039 (0.037, 0.041) \\
        Job income    & 0.610 (0.601, 0.619) & 0.905 (0.900, 0.910) & 0.286 (0.276, 0.298) \\
        Social        & 0.542 (0.528, 0.556) & 0.539 (0.524, 0.553) & 0.545 (0.530, 0.561) \\
        Family income & 0.913 (0.908, 0.919) & 0.979 (0.977, 0.980) & 0.842 (0.832, 0.851) \\
        \bottomrule
    \end{tabular}
\end{table}

For models with two mixture components, the marginal register probability is
$$Pr(Y = k) = \pi Pr(Y = k | g = 1) + (1-\pi)Pr(Y = k | g = 2)$$
where $\pi$ is the estimated mixing proportion.

For each BLB resample, the above quantities are computed, yielding bootstrap realisations of the marginal probabilities. The reported estimates and confidence intervals in Section 3.3 are obtained by averaging across BLB resamples.

\subsection*{Population Size and Overcoverage}

\begin{figure}[H]
\centering
\begin{subfigure}{0.5\textwidth}
  \centering
  \includegraphics[width=\linewidth]{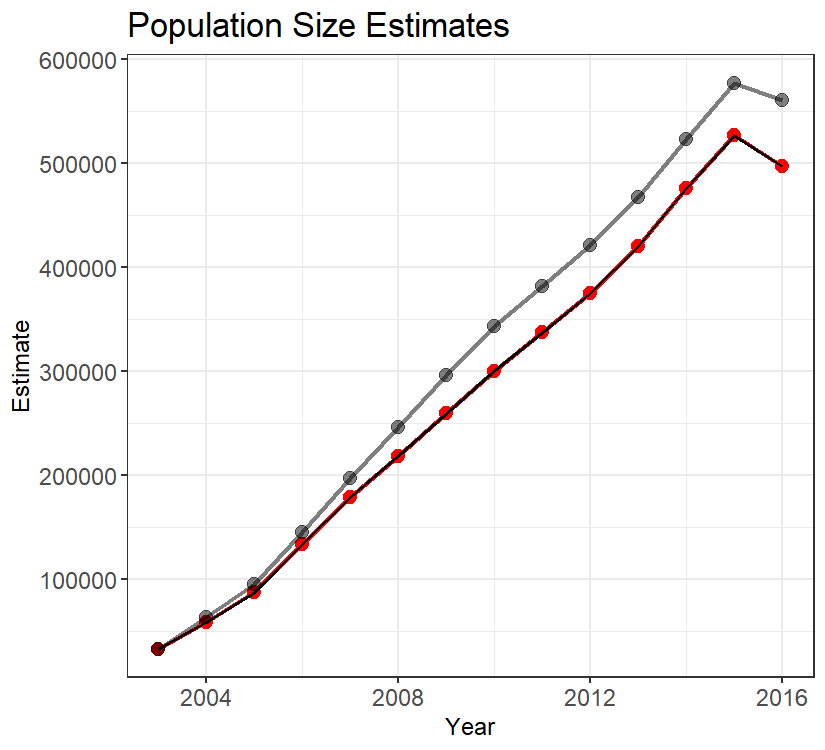}
  \caption{Population size estimates.}
  \label{pop_size}
\end{subfigure}%
\begin{subfigure}{0.5\textwidth}
  \centering
  \includegraphics[width=\linewidth]{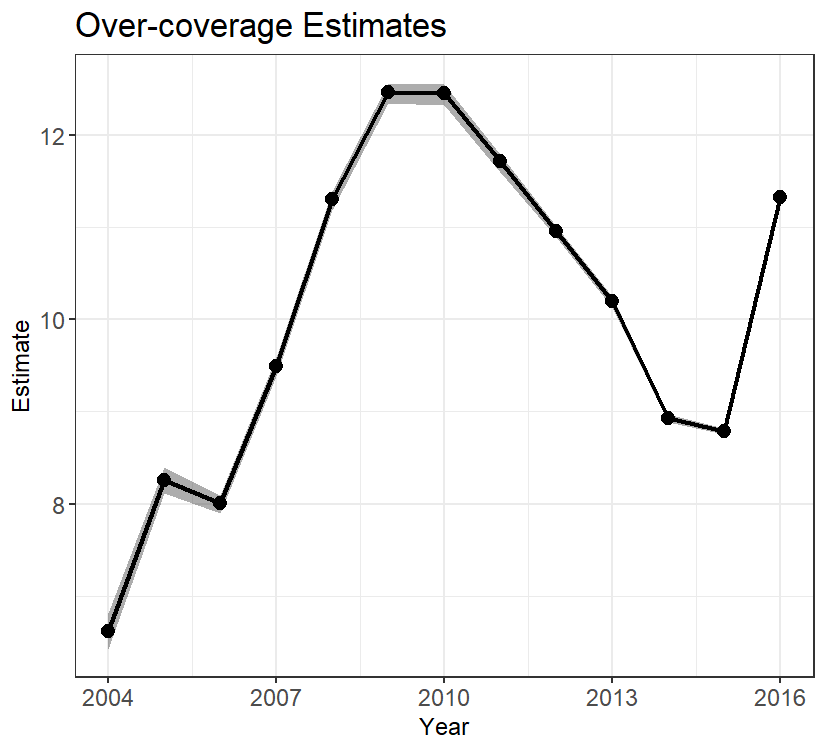}
  \caption{Overcoverage estimates.}
  \label{OC}
\end{subfigure}
\caption{Population size (plotted in red) and overcoverage estimates for the full population plotted alongside their $95\%$ confidence intervals. The grey line in Figure 6(a) shows the size of the RTB, i.e. the number of registered individuals in the country each year.}
\label{pop_and_OC}
\end{figure}

\begin{table}[H]
    \centering
    \caption{Population size and overcoverage estimates for the full population, alongside their 95\% confidence intervals.}
    \begin{tabular}{@{} c cc @{}}
        \toprule
        Year & Population Size (95\% CI) & Overcoverage (95\% CI) \\
        \midrule
        2004 & 58,466.61 (58,341.79, 58,588.07) & 6.620 (6.425, 6.819) \\
        2005 & 86,626.94 (86,493.97, 86,763.82) & 8.257 (8.112, 8.397) \\
        2006 & 133,375.28 (133,247.73, 133,534.16) & 8.003 (7.894, 8.091) \\
        2007 & 178,416.92 (178,232.08, 178,645.35) & 9.497 (9.381, 9.591) \\
        2008 & 218,317.14 (218,101.11, 218,607.73) & 11.306 (11.188, 11.394) \\
        2009 & 258,998.54 (258,713.61, 259,372.40) & 12.468 (12.341, 12.564) \\
        2010 & 300,178.39 (299,843.06, 300,607.61) & 12.452 (12.327, 12.550) \\
        2011 & 336,916.58 (336,609.81, 337,376.23) & 11.716 (11.596, 11.797) \\
        2012 & 375,098.63 (374,869.97, 375,360.08) & 10.955 (10.893, 11.010) \\
        2013 & 419,979.87 (419,698.63, 420,254.09) & 10.192 (10.134, 10.252) \\
        2014 & 476,111.61 (475,901.83, 476,331.60) & 8.929 (8.886, 8.969) \\
        2015 & 526,545.20 (526,344.76, 526,711.57) & 8.789 (8.760, 8.824) \\
        2016 & 497,418.41 (497,264.56, 497,466.29) & 11.328 (11.319, 11.355) \\
        \bottomrule
    \end{tabular}
\end{table}

\begin{figure}[H]
    \centering
  \includegraphics[width=\linewidth]{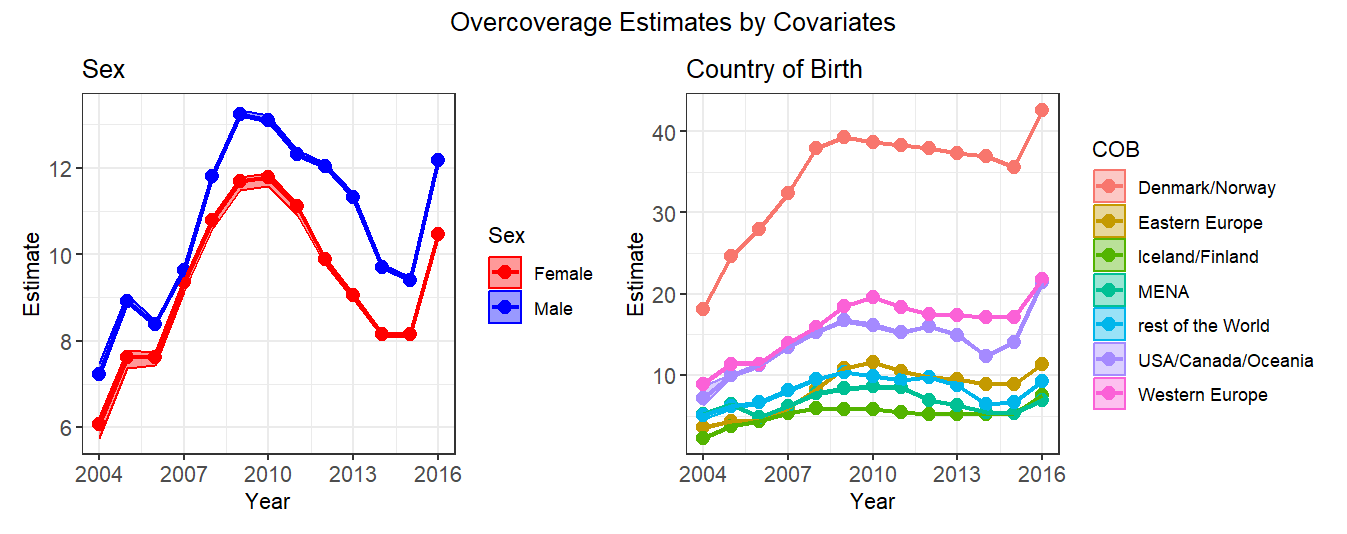}
    \caption{Overcoverage estimates with $95\%$ confidence intervlas for each year, decomposed by sex and country of birth covariates.}
    \label{OC_sex_cob}
\end{figure}
 
\begin{table}[H]
\centering
\caption{Population size estimates and overcoverage estimates with 95\% confidence intervals for each year, by the sex covariate.}
\begin{tabular}{@{} c l cc @{}}
\toprule
Year & Sex & Population Size (95\% CI) & Overcoverage (95\% CI) \\
\midrule
2004 & Female & 30,597.70 (30,539.55, 30,698.41) & 6.058 (5.749, 6.237) \\
2005 & Female & 44,728.03 (44,655.46, 44,849.48) & 7.621 (7.370, 7.771) \\
2006 & Female & 66,828.44 (66,743.71, 66,965.48) & 7.619 (7.430, 7.736) \\
2007 & Female & 87,457.57 (87,364.03, 87,653.54) & 9.349 (9.146, 9.446) \\
2008 & Female & 108,185.45 (108,075.46, 108,424.12) & 10.787 (10.590, 10.877) \\
2009 & Female & 129,458.89 (129,320.50, 129,774.10) & 11.692 (11.477, 11.786) \\
2010 & Female & 149,337.09 (149,176.38, 149,695.35) & 11.789 (11.577, 11.884) \\
2011 & Female & 168,112.22 (167,982.32, 168,520.19) & 11.113 (10.898, 11.182) \\
2012 & Female & 189,587.86 (189,491.80, 189,801.26) & 9.879 (9.777, 9.925) \\
2013 & Female & 212,501.81 (212,419.02, 212,716.36) & 9.052 (8.960, 9.087) \\
2014 & Female & 237,565.94 (237,502.08, 237,737.28) & 8.141 (8.075, 8.166) \\
2015 & Female & 260,145.05 (260,096.39, 260,263.94) & 8.142 (8.101, 8.160) \\
2016 & Female & 247,183.20 (247,031.78, 247,226.42) & 10.457 (10.441, 10.512) \\
\midrule
2004 & Male & 27,868.91 (27,792.38, 27,898.78) & 7.227 (7.128, 7.482) \\
2005 & Male & 41,898.92 (41,827.51, 41,923.32) & 8.925 (8.872, 9.081) \\
2006 & Male & 66,546.84 (66,491.70, 66,576.05) & 8.386 (8.345, 8.462) \\
2007 & Male & 90,959.35 (90,853.21, 90,999.23) & 9.640 (9.600, 9.745) \\
2008 & Male & 110,131.69 (110,010.20, 110,188.96) & 11.811 (11.765, 11.908) \\
2009 & Male & 129,539.66 (129,375.48, 129,606.48) & 13.230 (13.185, 13.339) \\
2010 & Male & 150,841.30 (150,644.75, 150,926.24) & 13.098 (13.049, 13.212) \\
2011 & Male & 168,804.36 (168,614.55, 168,874.78) & 12.309 (12.272, 12.407) \\
2012 & Male & 185,510.76 (185,367.12, 185,568.70) & 12.029 (12.002, 12.097) \\
2013 & Male & 207,478.06 (207,264.52, 207,553.50) & 11.331 (11.299, 11.422) \\
2014 & Male & 238,545.67 (238,386.85, 238,607.01) & 9.700 (9.676, 9.760) \\
2015 & Male & 266,400.16 (266,241.86, 266,462.58) & 9.412 (9.391, 9.466) \\
2016 & Male & 250,235.20 (250,225.59, 250,248.17) & 12.171 (12.167, 12.175) \\
\bottomrule
\end{tabular}
\end{table}
 
\begin{table}[H]
\centering
\caption{Population size estimates and overcoverage estimates with 95\% confidence intervals for each year, by the country of birth covariate.}
\begin{tabular}{@{} c l cc @{}}
\toprule
Year & Country of Birth & Population (95\% CI) & Overcoverage (95\% CI) \\
\midrule
2004 & Denmark/Norway & 5,662.00 (5,662.00, 5,662.00) & 18.084 (18.084, 18.084) \\
2005 & Denmark/Norway & 7,088.49 (7,087.27, 7,094.49) & 24.631 (24.567, 24.644) \\
2006 & Denmark/Norway & 8,862.63 (8,862.00, 8,870.70) & 27.911 (27.845, 27.916) \\
2007 & Denmark/Norway & 9,986.94 (9,986.00, 9,998.58) & 32.333 (32.254, 32.340) \\
2008 & Denmark/Norway & 9,971.40 (9,968.35, 9,989.42) & 37.970 (37.857, 37.989) \\
2009 & Denmark/Norway & 10,246.05 (10,240.84, 10,263.47) & 39.297 (39.194, 39.328) \\
2010 & Denmark/Norway & 10,680.54 (10,671.75, 10,693.72) & 38.688 (38.612, 38.739) \\
2011 & Denmark/Norway & 11,091.33 (11,080.27, 11,103.85) & 38.337 (38.267, 38.398) \\
2012 & Denmark/Norway & 11,446.66 (11,436.52, 11,464.35) & 37.871 (37.775, 37.926) \\
2013 & Denmark/Norway & 11,593.43 (11,579.19, 11,607.77) & 37.367 (37.289, 37.444) \\
2014 & Denmark/Norway & 11,781.06 (11,769.09, 11,797.85) & 36.946 (36.856, 37.010) \\
2015 & Denmark/Norway & 12,239.26 (12,229.98, 12,251.09) & 35.617 (35.554, 35.666) \\
2016 & Denmark/Norway & 9,961.35 (9,959.87, 9,969.13) & 42.622 (42.577, 42.631) \\
\midrule
2004 & Eastern Europe & 12,774.69 (12,751.11, 12,802.26) & 3.536 (3.328, 3.714) \\
2005 & Eastern Europe & 20,623.63 (20,599.16, 20,649.77) & 4.312 (4.191, 4.426) \\
2006 & Eastern Europe & 33,617.46 (33,583.15, 33,654.21) & 4.428 (4.324, 4.526) \\
2007 & Eastern Europe & 46,809.34 (46,771.31, 46,892.17) & 5.943 (5.777, 6.019) \\
2008 & Eastern Europe & 58,026.98 (57,981.00, 58,100.21) & 8.326 (8.210, 8.399) \\
2009 & Eastern Europe & 66,992.51 (66,937.45, 67,100.66) & 10.892 (10.748, 10.965) \\
2010 & Eastern Europe & 76,073.59 (76,017.91, 76,181.58) & 11.640 (11.515, 11.705) \\
2011 & Eastern Europe & 86,005.93 (85,973.12, 86,105.15) & 10.514 (10.411, 10.548) \\
2012 & Eastern Europe & 95,897.94 (95,870.64, 95,955.02) & 9.763 (9.709, 9.788) \\
2013 & Eastern Europe & 105,467.25 (105,441.43, 105,518.52) & 9.466 (9.421, 9.488) \\
2014 & Eastern Europe & 116,778.35 (116,748.92, 116,835.04) & 8.914 (8.870, 8.937) \\
2015 & Eastern Europe & 127,696.63 (127,679.98, 127,729.85) & 8.826 (8.802, 8.838) \\
2016 & Eastern Europe & 121,113.92 (121,107.69, 121,124.23) & 11.294 (11.286, 11.299) \\
\midrule
2004 & Iceland/Finland & 2,522.04 (2,522.00, 2,522.60) & 2.285 (2.263, 2.286) \\
2005 & Iceland/Finland & 3,103.20 (3,103.00, 3,105.38) & 3.687 (3.619, 3.693) \\
2006 & Iceland/Finland & 3,675.73 (3,674.00, 3,677.91) & 4.377 (4.321, 4.422) \\
2007 & Iceland/Finland & 4,212.11 (4,212.00, 4,213.61) & 5.282 (5.248, 5.284) \\
2008 & Iceland/Finland & 4,552.18 (4,552.00, 4,553.66) & 5.985 (5.955, 5.989) \\
2009 & Iceland/Finland & 5,096.79 (5,096.00, 5,097.48) & 5.824 (5.812, 5.839) \\
2010 & Iceland/Finland & 5,662.53 (5,660.14, 5,663.78) & 5.766 (5.745, 5.806) \\
2011 & Iceland/Finland & 6,252.03 (6,250.96, 6,253.36) & 5.401 (5.381, 5.417) \\
2012 & Iceland/Finland & 6,889.06 (6,888.52, 6,890.05) & 5.214 (5.200, 5.221) \\
2013 & Iceland/Finland & 7,411.08 (7,410.65, 7,412.26) & 5.156 (5.141, 5.162) \\
2014 & Iceland/Finland & 8,070.14 (8,069.00, 8,072.86) & 5.146 (5.115, 5.160) \\
2015 & Iceland/Finland & 9,051.11 (9,050.00, 9,053.33) & 5.284 (5.260, 5.295) \\
2016 & Iceland/Finland & 7,980.82 (7,979.00, 7,982.77) & 7.490 (7.468, 7.511) \\
\midrule
2004 & MENA & 13,925.82 (13,884.11, 13,935.69) & 5.189 (5.122, 5.473) \\
2005 & MENA & 20,280.87 (20,238.96, 20,305.49) & 6.397 (6.284, 6.591) \\
2006 & MENA & 37,207.27 (37,173.70, 37,257.07) & 4.877 (4.750, 4.963) \\
2007 & MENA & 53,061.57 (52,974.22, 53,147.60) & 6.195 (6.043, 6.350) \\
2008 & MENA & 65,484.90 (65,378.27, 65,639.90) & 7.759 (7.540, 7.909) \\
2009 & MENA & 80,114.52 (79,993.33, 80,281.98) & 8.376 (8.184, 8.514) \\
2010 & MENA & 94,236.32 (94,089.85, 94,452.48) & 8.645 (8.435, 8.787) \\
2011 & MENA & 105,936.58 (105,791.49, 106,219.22) & 8.501 (8.257, 8.626) \\
2012 & MENA & 121,201.02 (121,106.93, 121,302.80) & 6.967 (6.889, 7.039) \\
2013 & MENA & 140,569.27 (140,424.70, 140,689.82) & 6.255 (6.174, 6.351) \\
2014 & MENA & 165,778.71 (165,669.95, 165,869.28) & 5.429 (5.378, 5.491) \\
2015 & MENA & 190,590.77 (190,469.85, 190,679.25) & 5.318 (5.274, 5.378) \\
2016 & MENA & 185,410.84 (185,261.58, 185,440.16) & 6.911 (6.896, 6.986) \\
\bottomrule
\end{tabular}
\end{table}
 
\begin{table}[H]
\centering
\caption{(Continued) Population size estimates and overcoverage estimates with 95\% confidence intervals for each year, by the country of birth covariate.}
\begin{tabular}{@{} c l cc @{}}
\toprule
Year & Country of Birth & Population (95\% CI) & Overcoverage (95\% CI) \\
\midrule
2004 & USA/Canada/Oceania & 1,974.31 (1,942.45, 1,992.43) & 7.179 (6.327, 8.677) \\
2005 & USA/Canada/Oceania & 2,611.74 (2,605.44, 2,617.27) & 9.940 (9.749, 10.157) \\
2006 & USA/Canada/Oceania & 3,139.98 (3,130.97, 3,148.85) & 11.225 (10.974, 11.479) \\
2007 & USA/Canada/Oceania & 3,741.01 (3,734.42, 3,748.55) & 13.463 (13.288, 13.615) \\
2008 & USA/Canada/Oceania & 4,456.87 (4,452.29, 4,464.90) & 15.333 (15.180, 15.420) \\
2009 & USA/Canada/Oceania & 5,214.40 (5,201.61, 5,225.50) & 16.703 (16.526, 16.907) \\
2010 & USA/Canada/Oceania & 6,203.77 (6,183.02, 6,218.42) & 16.109 (15.910, 16.389) \\
2011 & USA/Canada/Oceania & 7,008.66 (6,991.72, 7,025.45) & 15.252 (15.049, 15.457) \\
2012 & USA/Canada/Oceania & 7,518.77 (7,505.72, 7,530.28) & 15.973 (15.844, 16.118) \\
2013 & USA/Canada/Oceania & 8,394.19 (8,376.25, 8,408.38) & 14.970 (14.826, 15.151) \\
2014 & USA/Canada/Oceania & 9,227.33 (9,217.69, 9,234.56) & 12.371 (12.302, 12.463) \\
2015 & USA/Canada/Oceania & 9,165.45 (9,157.84, 9,172.49) & 14.109 (14.043, 14.180) \\
2016 & USA/Canada/Oceania & 7,690.86 (7,690.00, 7,693.01) & 21.482 (21.460, 21.491) \\
\midrule
2004 & Western Europe & 6,671.29 (6,656.03, 6,682.05) & 8.875 (8.728, 9.083) \\
2005 & Western Europe & 9,585.40 (9,559.81, 9,599.45) & 11.402 (11.272, 11.639) \\
2006 & Western Europe & 13,879.70 (13,856.64, 13,893.80) & 11.397 (11.307, 11.544) \\
2007 & Western Europe & 18,080.81 (18,052.04, 18,101.56) & 13.938 (13.839, 14.075) \\
2008 & Western Europe & 22,374.35 (22,342.32, 22,395.44) & 15.845 (15.765, 15.965) \\
2009 & Western Europe & 25,043.03 (25,006.20, 25,079.97) & 18.506 (18.386, 18.626) \\
2010 & Western Europe & 27,438.28 (27,395.45, 27,479.36) & 19.642 (19.522, 19.767) \\
2011 & Western Europe & 30,920.84 (30,880.33, 30,949.73) & 18.369 (18.293, 18.476) \\
2012 & Western Europe & 34,456.62 (34,417.36, 34,482.25) & 17.536 (17.475, 17.630) \\
2013 & Western Europe & 38,254.42 (38,211.30, 38,284.65) & 17.432 (17.367, 17.525) \\
2014 & Western Europe & 42,382.68 (42,344.70, 42,416.66) & 17.090 (17.024, 17.164) \\
2015 & Western Europe & 46,121.58 (46,084.82, 46,142.43) & 17.104 (17.067, 17.170) \\
2016 & Western Europe & 41,093.71 (41,092.29, 41,097.00) & 21.810 (21.803, 21.812) \\
\midrule
2004 & Rest of the World & 14,936.47 (14,902.57, 15,015.18) & 5.099 (4.599, 5.314) \\
2005 & Rest of the World & 23,333.61 (23,275.29, 23,419.67) & 6.129 (5.782, 6.363) \\
2006 & Rest of the World & 32,992.51 (32,938.94, 33,071.81) & 6.664 (6.439, 6.815) \\
2007 & Rest of the World & 42,525.14 (42,466.37, 42,600.34) & 8.092 (7.929, 8.219) \\
2008 & Rest of the World & 53,450.45 (53,387.94, 53,521.94) & 9.542 (9.421, 9.648) \\
2009 & Rest of the World & 66,291.25 (66,195.26, 66,409.39) & 10.404 (10.244, 10.534) \\
2010 & Rest of the World & 79,883.37 (79,772.59, 80,003.59) & 9.893 (9.757, 10.018) \\
2011 & Rest of the World & 89,701.21 (89,598.04, 89,788.15) & 9.387 (9.299, 9.491) \\
2012 & Rest of the World & 97,688.55 (97,604.08, 97,794.02) & 9.777 (9.679, 9.855) \\
2013 & Rest of the World & 108,290.23 (108,206.49, 108,383.83) & 8.750 (8.671, 8.820) \\
2014 & Rest of the World & 122,093.35 (122,050.92, 122,145.12) & 6.402 (6.363, 6.435) \\
2015 & Rest of the World & 131,680.39 (131,648.50, 131,712.97) & 6.645 (6.622, 6.668) \\
2016 & Rest of the World & 124,166.91 (124,163.16, 124,168.87) & 9.310 (9.308, 9.312) \\
\bottomrule
\end{tabular}
\end{table}
 
\begin{table}[H]
\centering
\small
\caption{Overcoverage estimates with 95\% confidence intervals for the proposed model and three restricted versions: without the finite-mixture model, without false-positives, and without both.}
\begin{tabular}{@{} c cccc @{}}
\toprule
Year & New Model & No FMM & No FP & No FMM or FP \\
\midrule
2004 & 6.619 (6.425, 6.819) & 7.117 (7.117, 7.117) & 4.419 (4.396, 4.458) & 4.709 (4.651, 4.768) \\
2005 & 8.257 (8.112, 8.397) & 8.674 (8.628, 8.748) & 5.629 (5.593, 5.667) & 5.868 (5.820, 5.912) \\
2006 & 8.003 (7.894, 8.091) & 8.258 (8.221, 8.304) & 5.543 (5.513, 5.579) & 5.673 (5.640, 5.699) \\
2007 & 9.497 (9.381, 9.591) & 9.823 (9.788, 9.863) & 6.365 (6.336, 6.397) & 6.485 (6.451, 6.522) \\
2008 & 11.306 (11.188, 11.394) & 11.677 (11.630, 11.743) & 7.667 (7.631, 7.715) & 7.820 (7.786, 7.849) \\
2009 & 12.468 (12.341, 12.564) & 12.878 (12.833, 12.968) & 8.618 (8.574, 8.674) & 8.797 (8.749, 8.831) \\
2010 & 12.452 (12.327, 12.550) & 12.894 (12.843, 12.974) & 8.606 (8.559, 8.665) & 8.781 (8.731, 8.819) \\
2011 & 11.716 (11.596, 11.797) & 12.096 (12.053, 12.171) & 7.994 (7.955, 8.037) & 8.075 (8.048, 8.108) \\
2012 & 10.955 (10.893, 11.010) & 11.205 (11.170, 11.267) & 7.868 (7.840, 7.898) & 7.943 (7.922, 7.966) \\
2013 & 10.192 (10.134, 10.252) & 10.443 (10.411, 10.508) & 7.222 (7.197, 7.248) & 7.268 (7.251, 7.293) \\
2014 & 8.929 (8.886, 8.969) & 9.111 (9.079, 9.149) & 6.030 (6.014, 6.047) & 6.046 (6.036, 6.056) \\
2015 & 8.789 (8.760, 8.824) & 8.927 (8.904, 8.954) & 5.903 (5.886, 5.916) & 5.915 (5.907, 5.922) \\
2016 & 11.328 (11.319, 11.355) & 11.489 (11.414, 11.514) & 7.745 (7.724, 7.775) & 7.804 (7.802, 7.804) \\
\bottomrule
\end{tabular}
\end{table}

\begin{figure}[H]
    \centering
    \includegraphics[width=\textwidth]{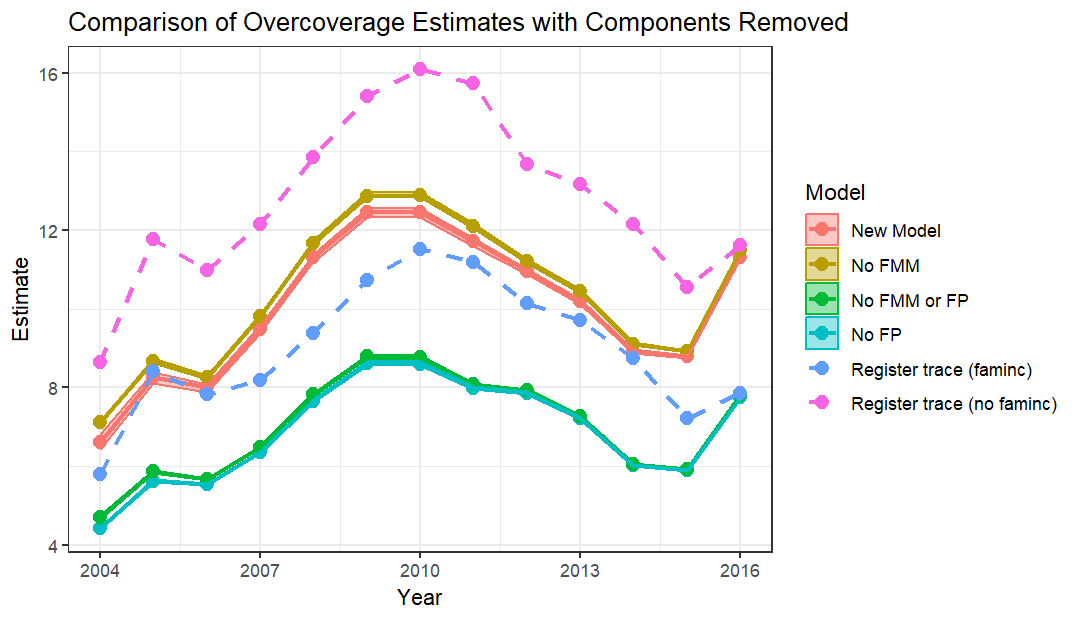}
    \caption{Overcoverage estimates for reduced model variants compared with the full model. The estimates denoted by ``New Model" are from the model proposed in this paper. The other three models are obtained by removing different aspects: the finite mixture model, accounting for false positives, and both at the same time. All estimates are plotted alongside their corresponding $95\%$ confidence intervals. Also plotted are the overcoverage estimates obtained by the existing register-trace appraoch, plotted with and without the family income register.}
\end{figure}

\begin{table}[H]
\centering
\caption{Overcoverage estimates with 95\% confidence intervals for the proposed model and the register trace approach (with and without the family income register).}
\begin{tabular}{@{} c ccc @{}}
\toprule
Year & New Model & RT (with) & RT (without) \\
\midrule
2004 & 6.619 (6.425, 6.819) & 5.793 & 8.644 \\
2005 & 8.257 (8.112, 8.397) & 8.412 & 11.770 \\
2006 & 8.003 (7.894, 8.091) & 7.836 & 10.979 \\
2007 & 9.497 (9.381, 9.591) & 8.198 & 12.149 \\
2008 & 11.306 (11.188, 11.394) & 9.392 & 13.855 \\
2009 & 12.468 (12.341, 12.564) & 10.734 & 15.399 \\
2010 & 12.452 (12.327, 12.550) & 11.506 & 16.084 \\
2011 & 11.716 (11.596, 11.797) & 11.183 & 15.727 \\
2012 & 10.955 (10.893, 11.010) & 10.159 & 13.682 \\
2013 & 10.192 (10.134, 10.252) & 9.706 & 13.169 \\
2014 & 8.929 (8.886, 8.969) & 8.741 & 12.162 \\
2015 & 8.789 (8.760, 8.824) & 7.217 & 10.564 \\
2016 & 11.328 (11.319, 11.355) & 7.870 & 11.620 \\
\bottomrule
\end{tabular}
\end{table}

\subsection*{Uncertain Sightings}

\begin{figure}[H]
    \centering
    \includegraphics[width=\textwidth]{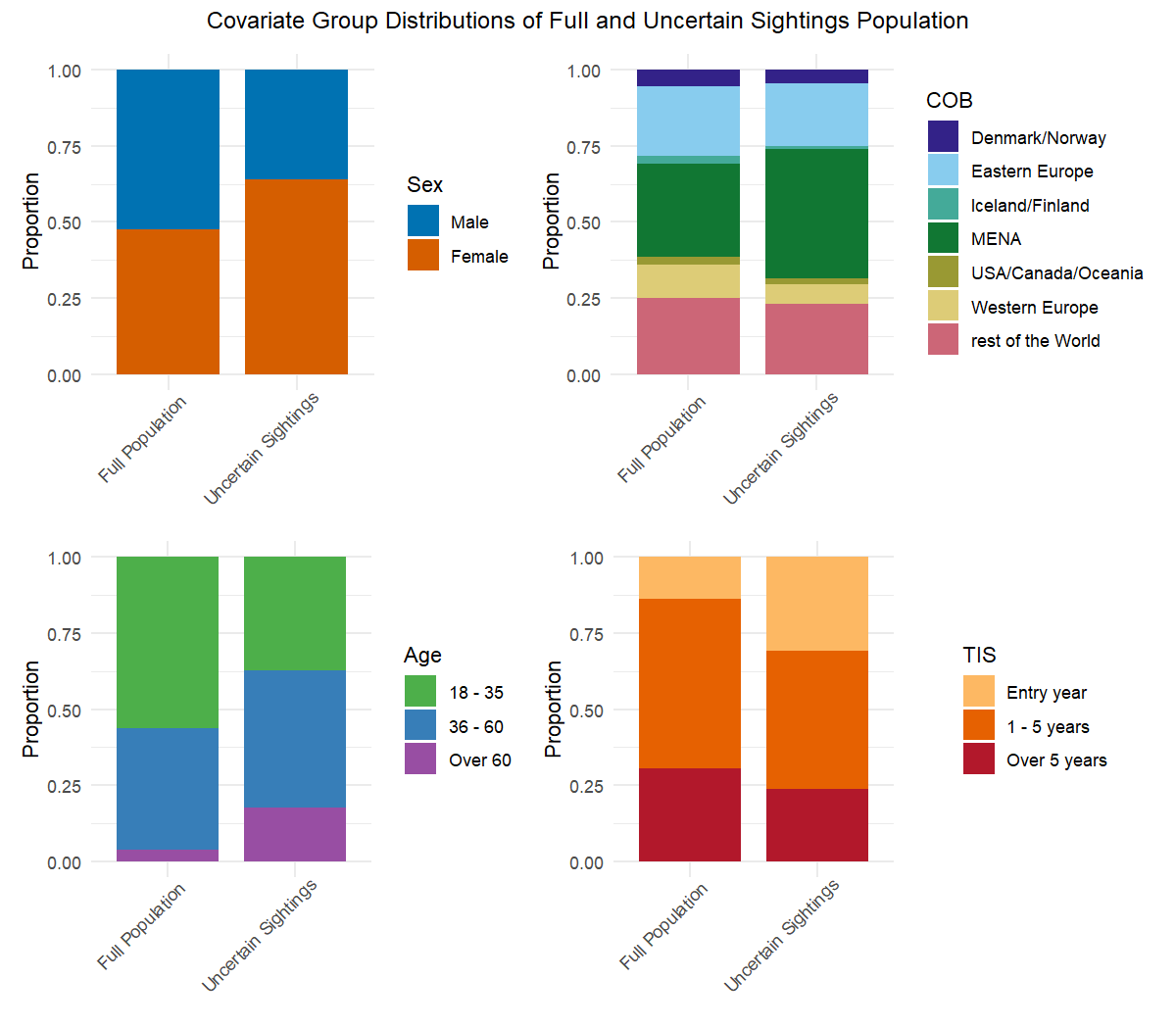}
    \caption{Proportion of individuals in each sex, country of birth group, age group and time since first entering Sweden group for different groups in the population. We compare the proportion of individuals in each group in the full population vs those observed only in the family income register at least one year, i.e. uncertain sightings. }
\end{figure}

\begin{figure}[H]
    \centering
    \includegraphics[width=\textwidth]{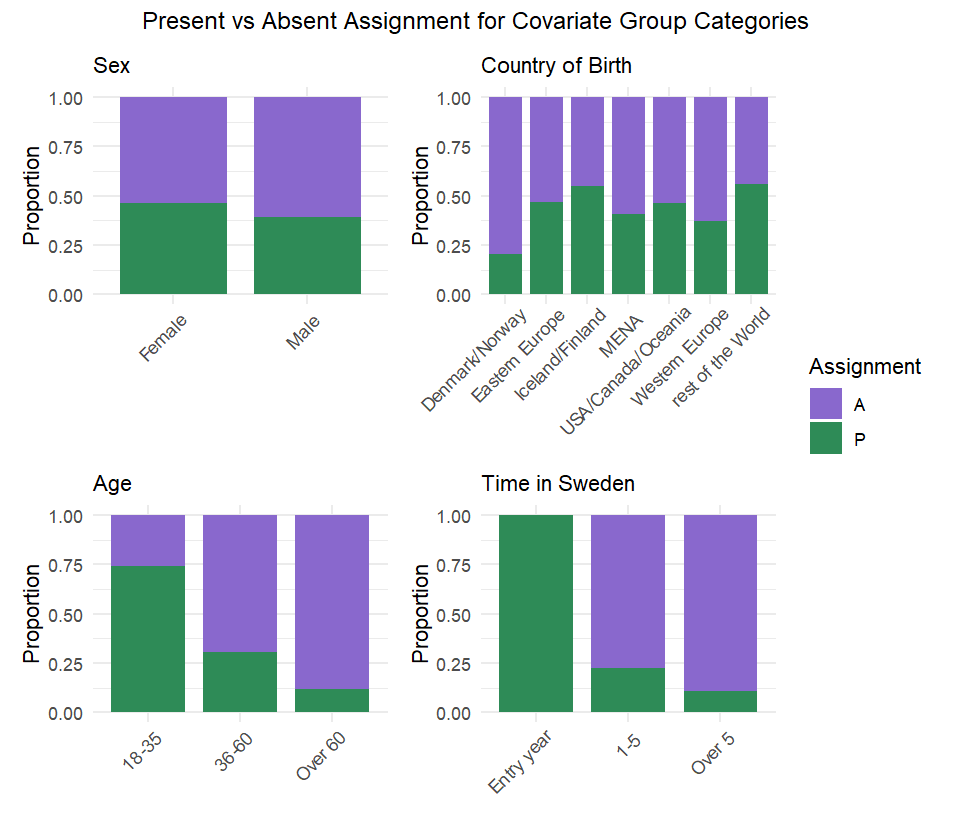}
    \caption{Proportion of present vs absent assignments for individuals who are uncertain sightings broken down by covariate category for sex, country of birth, age group and time since first entering Sweden group.}
\end{figure}

\begin{figure}[H]
    \centering
    \includegraphics[width=0.8\textwidth]{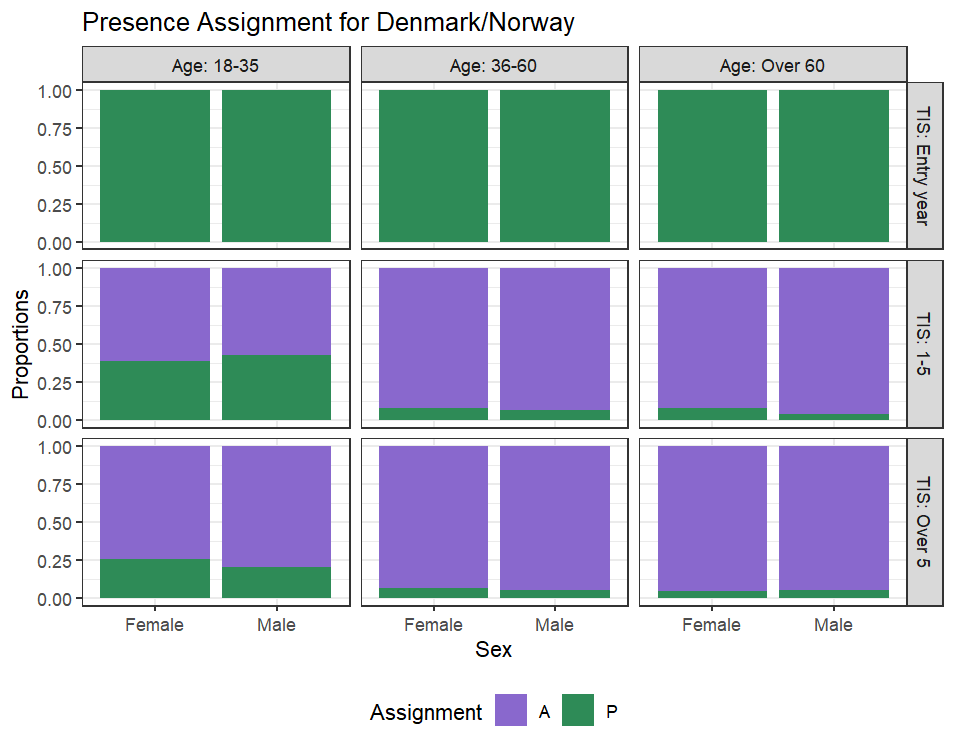}
    \caption{Proportion of present vs absent assignments for individuals who are uncertain sightings, conditional on country of birth group 1 - Denmark/Norway, broken down by covariate group combination.}
\end{figure}
\begin{figure}[H]
    \centering
    \includegraphics[width=0.8\textwidth]{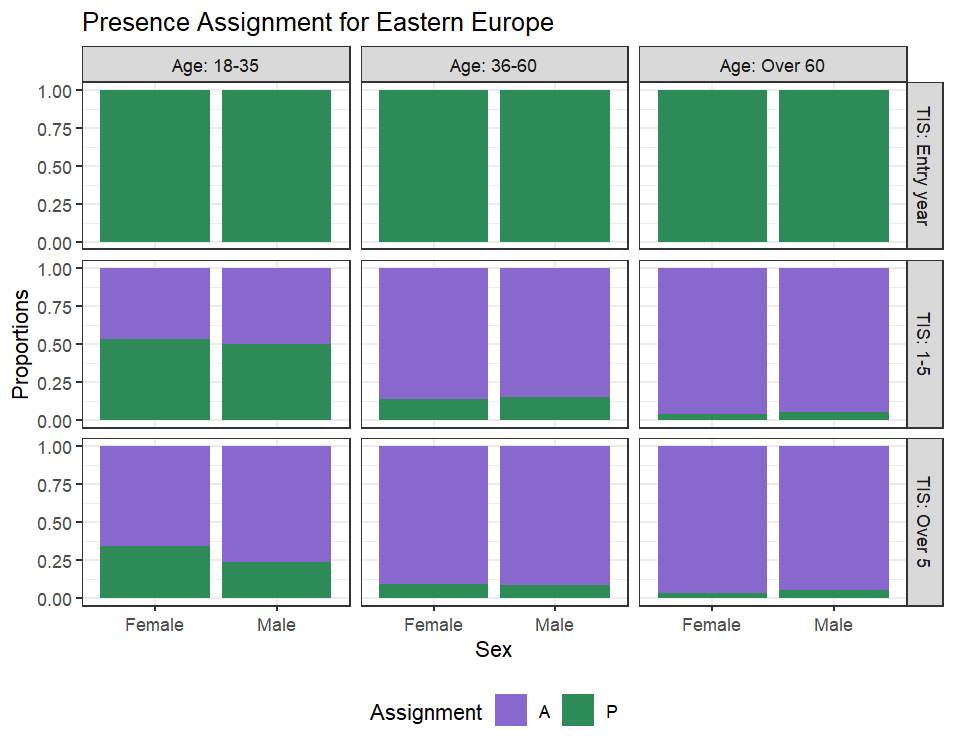}
    \caption{Proportion of present vs absent assignments for individuals who are uncertain sightings, conditional on country of birth group 2 - Eastern Europe, broken down by covariate group combination.}
\end{figure}
\begin{figure}[H]
    \centering
    \includegraphics[width=0.8\textwidth]{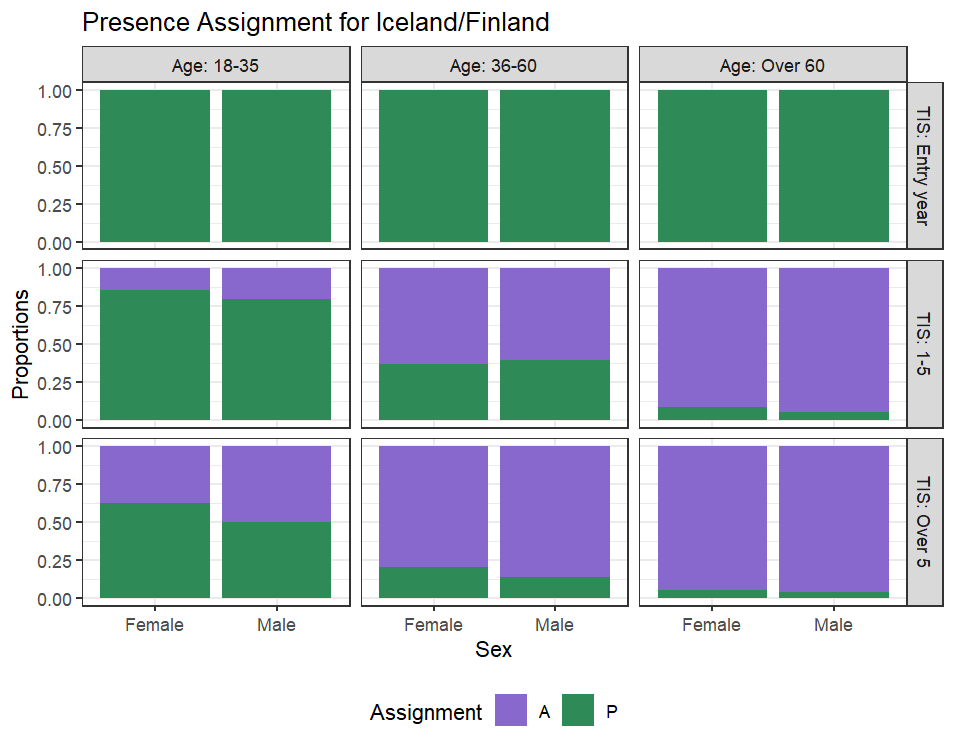}
    \caption{Proportion of present vs absent assignments for individuals who are uncertain sightings, conditional on country of birth group 3 - Iceland/Finland, broken down by covariate group combination.}
\end{figure}
\begin{figure}[H]
    \centering
    \includegraphics[width=0.8\textwidth]{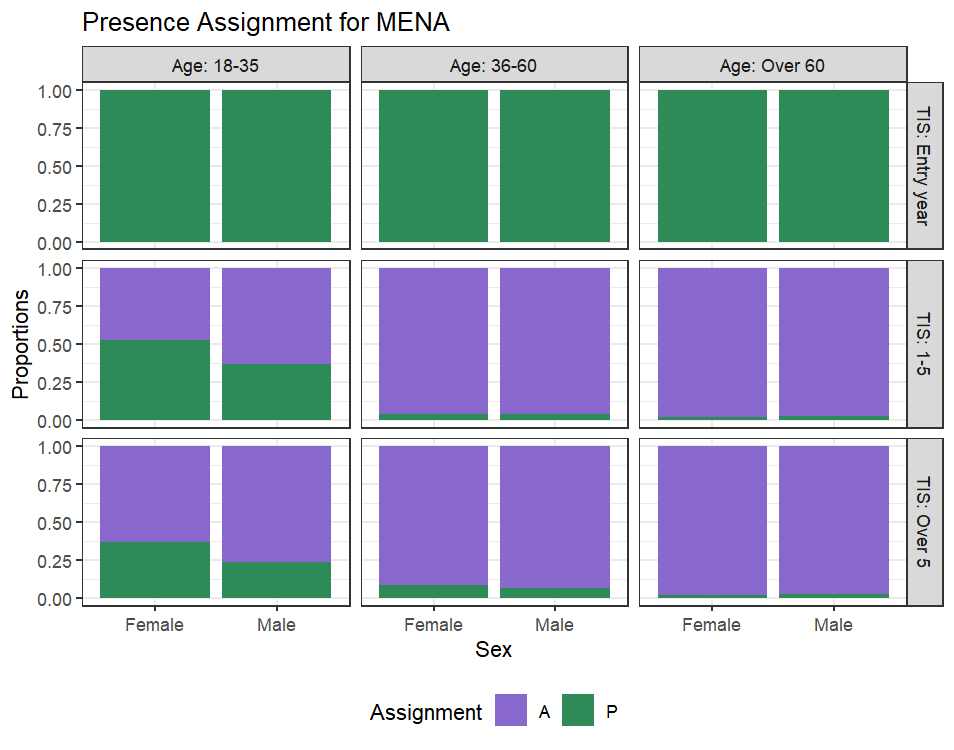}
    \caption{Proportion of present vs absent assignments for individuals who are uncertain sightings, conditional on country of birth group 4 - MENA, broken down by covariate group combination.}
\end{figure}
\begin{figure}[H]
    \centering
    \includegraphics[width=0.8\textwidth]{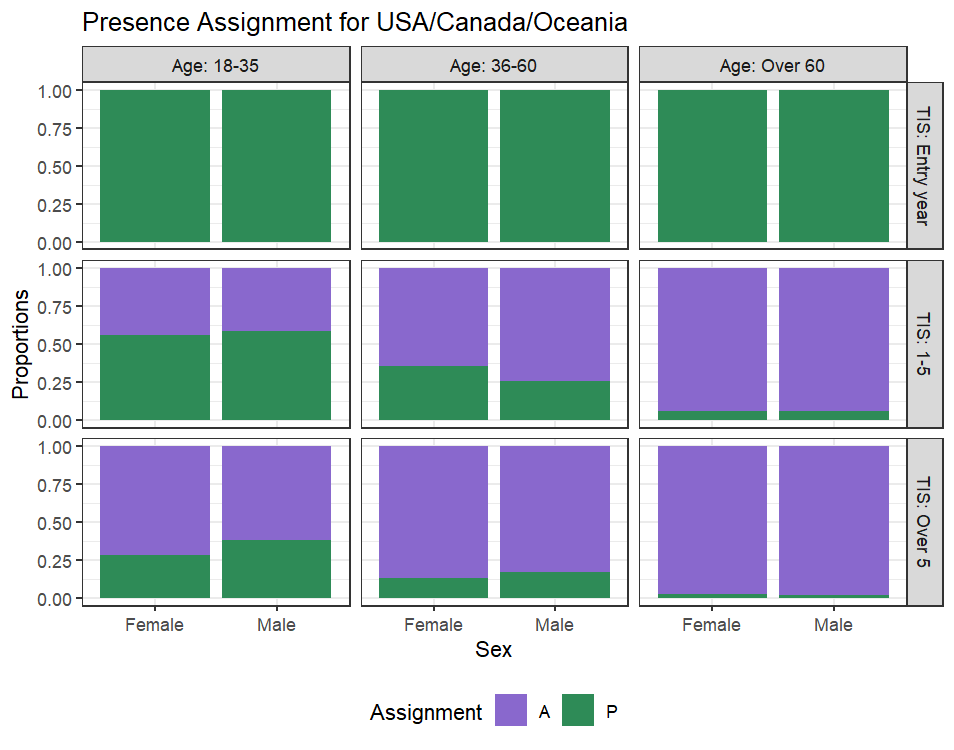}
    \caption{Proportion of present vs absent assignments for individuals who are uncertain sightings, conditional on country of birth group 5 - USA/Canada/Oceania, broken down by covariate group combination.}
\end{figure}
\begin{figure}[H]
    \centering
    \includegraphics[width=0.8\textwidth]{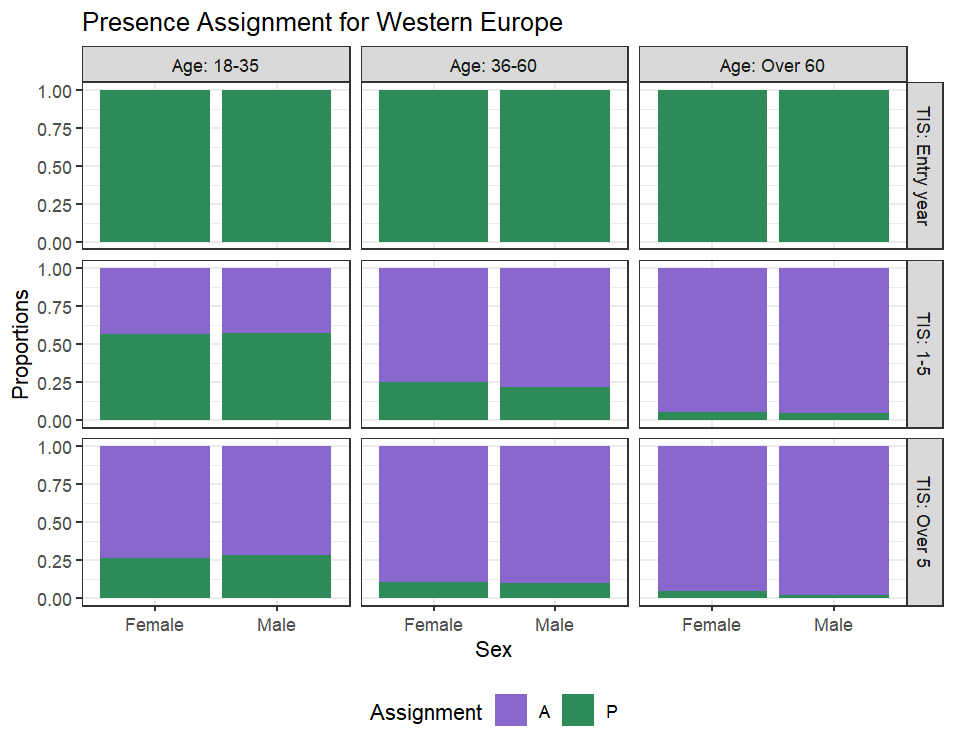}
    \caption{Proportion of present vs absent assignments for individuals who are uncertain sightings, conditional on country of birth group 6 - Western Europe, broken down by covariate group combination.}
\end{figure}
\begin{figure}[H]
    \centering
    \includegraphics[width=0.8\textwidth]{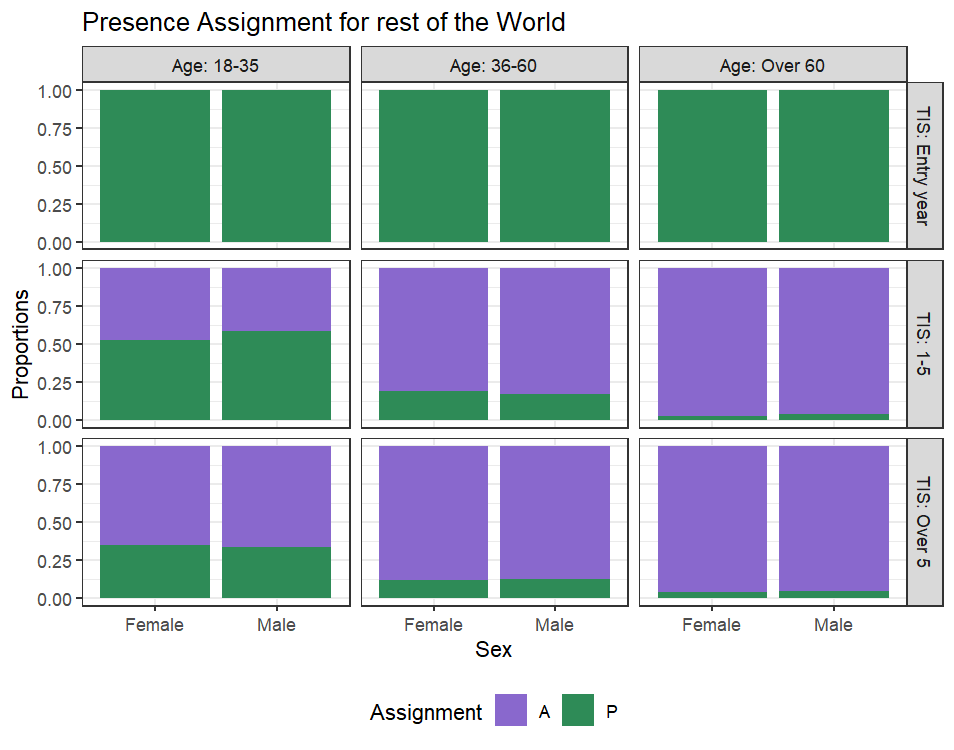}
    \caption{Proportion of present vs absent assignments for individuals who are uncertain sightings, conditional on country of birth group 7 - rest of the World, broken down by covariate group combination.}
\end{figure}

\begin{figure}[H]
    \centering
    \includegraphics[width=0.9\linewidth]{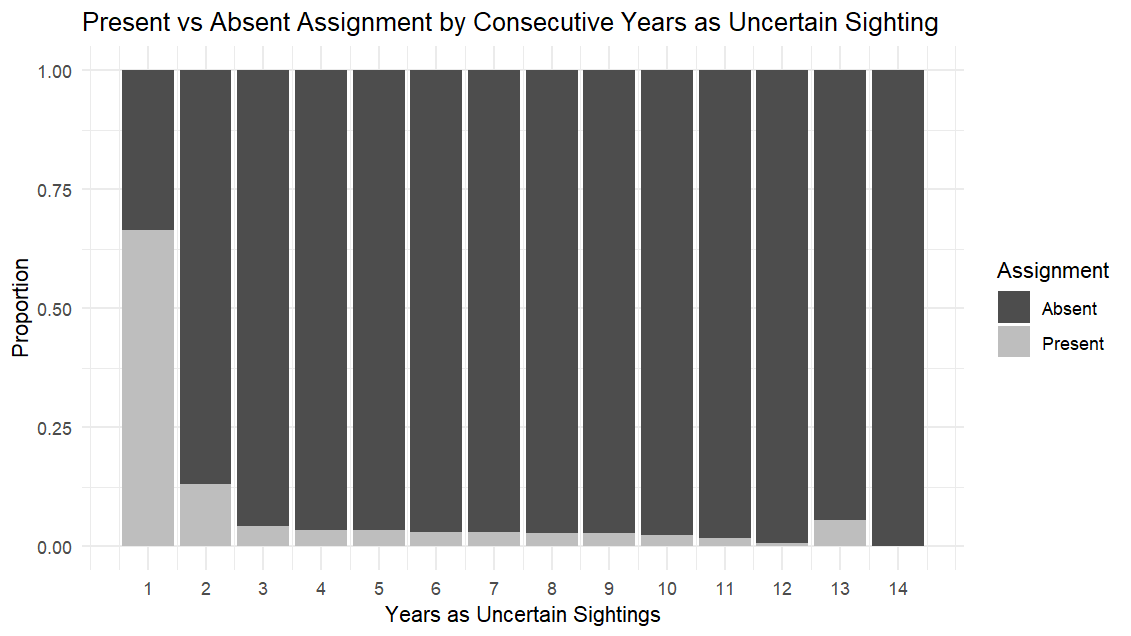}
    \caption{This figure highlights the proportion of individuals who are uncertain sightings that are assigned as present or absent in the county, conditional on the number of consecutive years they have been an uncertain sighting.}
\end{figure}

\begin{table}[H]
\centering
\caption{Estimated probability of true presence for individuals observed only in the family income register, as a function of consecutive years of such observations. These estimated probabilities are presented alongside their $95\%$ confidence intervals. The counts of individuals in each covariate group for each number of consecutive years should be taken into account - these counts are presented in Table \ref{tab:fp_counts}}
\label{tab:fp_means}
\footnotesize
\begin{tabular}{@{} l *{4}{r} @{}}
\toprule
& \multicolumn{4}{c}{Consecutive years in family income register only} \\
\cmidrule(lr){2-5}
Group & 1 & 2 & 3 & 4 \\
\midrule
\multicolumn{5}{l}{\textit{Sex}} \\
\quad Male   & 0.606 (0.6596, 0.611) & 0.111 (0.097, 0.115) & 0.044 (0.043, 0.045) & 0.035 (0.034, 0.035) \\
\quad Female & 0.697 (0.693, 0.714) & 0.143 (0.134, 0.159) & 0.040 (0.040, 0.041) & 0.033 (0.033, 0.033) \\
\midrule
\multicolumn{5}{l}{\textit{Country of birth}} \\
\quad Denmark/Norway     & 0.365 (0.361, 0.385) & 0.087 (0.084, 0.092) & 0.052 (0.052, 0.052) & 0.042 (0.042, 0.042) \\
\quad Eastern Europe     & 0.715 (0.710, 0.726) & 0.144 (0.128, 0.184) & 0.037 (0.037, 0.037) & 0.029 (0.029, 0.029) \\
\quad Iceland/Finland    & 0.794 (0.792, 0.798) & 0.333 (0.326, 0.348) & 0.109 (0.104, 0.124) & 0.061 (0.059, 0.078) \\
\quad MENA               & 0.630 (0.622, 0.651) & 0.097 (0.089, 0.098) & 0.023 (0.023, 0.023) & 0.025 (0.025, 0.025) \\
\quad USA/Canada/Oceania & 0.687 (0.685, 0.694) & 0.251 (0.248, 0.261) & 0.114 (0.113, 0.123) & 0.055 (0.055, 0.055) \\
\quad Western Europe     & 0.600 (0.581, 0.605) & 0.178 (0.139, 0.186) & 0.077 (0.074, 0.078) & 0.043 (0.043, 0.043) \\
\quad Rest of world      & 0.763 (0.761, 0.767) & 0.172 (0.162, 0.190) & 0.058 (0.058, 0.058) & 0.051 (0.051, 0.051) \\
\midrule
& \multicolumn{4}{c}{} \\[-1.5ex]
\cmidrule(lr){2-5}
Group & 5 & 6 & 7 & 8 \\
\midrule
\multicolumn{5}{l}{\textit{Sex}} \\
\quad Male   & 0.033 (0.032, 0.033) & 0.033 (0.033, 0.033) & 0.037 (0.037, 0.037) & 0.026 (0.026, 0.026) \\
\quad Female & 0.035 (0.035, 0.035) & 0.029 (0.029, 0.029) & 0.026 (0.026, 0.026) & 0.028 (0.028, 0.028) \\
\midrule
\multicolumn{5}{l}{\textit{Country of birth}} \\
\quad Denmark/Norway     & 0.043 (0.043, 0.043) & 0.037 (0.037, 0.037) & 0.040 (0.040, 0.040) & 0.027 (0.027, 0.027) \\
\quad Eastern Europe     & 0.043 (0.043, 0.043) & 0.025 (0.025, 0.025) & 0.032 (0.032, 0.032) & 0.016 (0.016, 0.016) \\
\quad Iceland/Finland    & 0.091 (0.091, 0.099) & 0.071 (0.071, 0.071) & 0.000 (0.000, 0.000) & 0.048 (0.048, 0.048) \\
\quad MENA               & 0.024 (0.024, 0.024) & 0.024 (0.024, 0.024) & 0.021 (0.021, 0.021) & 0.027 (0.027, 0.027) \\
\quad USA/Canada/Oceania & 0.021 (0.021, 0.021) & 0.048 (0.048, 0.048) & 0.016 (0.016, 0.016) & 0.023 (0.023, 0.023) \\
\quad Western Europe     & 0.043 (0.043, 0.043) & 0.036 (0.036, 0.036) & 0.025 (0.025, 0.025) & 0.042 (0.042, 0.042) \\
\quad Rest of world      & 0.037 (0.037, 0.037) & 0.052 (0.052, 0.052) & 0.063 (0.063, 0.063) & 0.038 (0.038, 0.038) \\
\midrule
& \multicolumn{4}{c}{} \\[-1.5ex]
\cmidrule(lr){2-5}
Group & 9 & 10 & 11 & 12 \\
\midrule
\multicolumn{5}{l}{\textit{Sex}} \\
\quad Male   & 0.035 (0.035, 0.035) & 0.031 (0.031, 0.031) & 0.022 (0.022, 0.022) & 0.016 (0.016, 0.016) \\
\quad Female & 0.025 (0.025 0.025) & 0.021 (0.021, 0.021) & 0.015 (0.015, 0.015) & 0.000 (0.000, 0.000) \\
\midrule
\multicolumn{5}{l}{\textit{Country of birth}} \\
\quad Denmark/Norway     & 0.025 (0.025, 0.025) & 0.000 (0.000, 0.000) & 0.025 (0.025, 0.025) & 0.045 (0.045, 0.045) \\
\quad Eastern Europe     & 0.031 (0.031, 0.031) & 0.064 (0.064, 0.064) & 0.036 (0.036, 0.036) & 0.000 (0.000, 0.000) \\
\quad Iceland/Finland    & 0.000 (0.000, 0.000) & 0.000 (0.000, 0.000) & 0.000 (0.000, 0.000) & 0.000 (0.000, 0.000) \\
\quad MENA               & 0.025 (0.025, 0.025) & 0.007 (0.007, 0.007) & 0.008 (0.008, 0.008) & 0.000 (0.000, 0.000) \\
\quad USA/Canada/Oceania & 0.000 (0.000, 0.000) & 0.000 (0.000, 0.000) & 0.000 (0.000, 0.000) & 0.000 (0.000, 0.000) \\
\quad Western Europe     & 0.039 (0.039, 0.039) & 0.040 (0.040, 0.040) & 0.000 (0.000, 0.000) & 0.000 (0.000, 0.000) \\
\quad Rest of world      & 0.044 (0.044, 0.044) & 0.024 (0.024, 0.024) & 0.25 (0.025, 0.025) & 0.000 (0.000, 0.000) \\
\midrule
& \multicolumn{4}{c}{} \\[-1.5ex]
\cmidrule(lr){2-5}
Group & 13 & 14 & & \\
\midrule
\multicolumn{5}{l}{\textit{Sex}} \\
\quad Male   & 0.136 (0.136, 0.136) & 0.000 (0.000, 0.000) \\
\quad Female & 0.000 (0.000, 0.000) & 0.000 (0.000, 0.000) \\
\midrule
\multicolumn{5}{l}{\textit{Country of birth}} \\
\quad Denmark/Norway     & 0.125 (0.125, 0.125) & 0.000 (0.000, 0.000) \\
\quad Eastern Europe     & 0.143 (0.143, 0.143) & 0.000 (0.000, 0.000) \\
\quad Iceland/Finland    & 0.000 (0.000, 0.000) & 0.000 (0.000, 0.000) \\
\quad MENA               & 0.000 (0.000, 0.000) & 0.000 (0.000, 0.000) \\
\quad USA/Canada/Oceania & 0.000 (0.000, 0.000) & 0.000 (0.000, 0.000) \\
\quad Western Europe     & 0.333 (0.333, 0.333) & 0.000 (0.000, 0.000) \\
\quad Rest of world      & 0.000 (0.000, 0.000) & 0.000 (0.000, 0.000) \\
\bottomrule
\end{tabular}
\end{table}

 \begin{table}[H]
\centering
\caption{Number of observations by covariate group and consecutive years observed in the family income register only.}
\label{tab:fp_counts}
\begin{tabular}{@{} l rrrrrrr @{}}
\toprule
& \multicolumn{7}{c}{Consecutive years in family income register only} \\
\cmidrule(lr){2-8}
Group & 1 & 2 & 3 & 4 & 5 & 6 & 7 \\
\midrule
\multicolumn{8}{l}{\textit{Sex}} \\
\quad Male   & 51,123 & 15,343 & 7,424 & 4,320 & 2,677 & 1,741 & 1,145 \\
\quad Female & 92,828 & 26,970 & 12,422 & 6,956 & 4,350 & 2,866 & 1,892 \\
\midrule
\multicolumn{8}{l}{\textit{Country of birth}} \\
\quad Denmark/Norway     &  6,986 &  3,270 & 1,773 & 1,080 &  673 &  409 &  277 \\
\quad Eastern Europe     & 29,244 &  8,451 & 4,015 & 2,321 & 1,489 & 1,009 &  659 \\
\quad Iceland/Finland    &  1,341 &    412 &  212 &  118 &   77 &   42 &   27 \\
\quad MENA               & 62,667 & 18,086 & 8,572 & 4,902 & 3,062 & 2,077 & 1,381 \\
\quad USA/Canada/Oceania &  2,768 &    908 &  416 &  236 &  144 &   84 &   63 \\
\quad Western Europe     &  9,012 &  3,371 & 1,738 & 1,036 &  652 &  411 &  279 \\
\quad Rest of world      & 31,933 &  7,815 & 3,120 & 1,583 &  930 &  575 &  351 \\
\midrule
& \multicolumn{7}{c}{} \\[-1.5ex]
\cmidrule(lr){2-8}
Group & 8 & 9 & 10 & 11 & 12 & 13 & 14 \\
\midrule
\multicolumn{8}{l}{\textit{Sex}} \\
\quad Male   & 726 & 462 & 262 & 134 & 64 & 22 & 5 \\
\quad Female & 1,187 & 747 & 429 & 202 & 93 & 33 & 12 \\
\midrule
\multicolumn{8}{l}{\textit{Country of birth}} \\
\quad Denmark/Norway     & 186 & 120 & 76 & 40 & 22 & 8 & 2 \\
\quad Eastern Europe     & 428 & 294 & 171 & 83 & 36 & 7 & 3 \\
\quad Iceland/Finland    & 21 & 8 & 4 & 3 & 1 & 1 & 1 \\
\quad MENA               & 853 & 518 & 284 & 132 & 50 & 22 & 5 \\
\quad USA/Canada/Oceania & 44 & 30 & 21 & 14 & 7 & 3 & 1 \\
\quad Western Europe     & 168 & 102 & 50 & 24 & 14 & 3 & 1 \\
\quad Rest of world      & 213 & 137 & 85 & 40 & 27 & 11 & 4 \\
\bottomrule
\end{tabular}
\end{table}